\documentclass[aps,pre,twocolumn,groupedaddress]{revtex4}
\usepackage{amssymb}
\usepackage{amsmath}
\usepackage{cases}
\usepackage{graphicx,color}

\definecolor{r}{rgb}{1,0,0}   
\definecolor{g}{rgb}{0,1,0}   
\definecolor{b}{rgb}{0,0,1}

\begin{document}
\title{Characterizing Pixel and Point Patterns with a Hyperuniformity Disorder Length}

\author{A.T. Chieco$^{1}$, R. Dreyfus$^2$ and D.J. Durian$^{1}$}
\affiliation{
	$^1$Department of Physics and Astronomy, University of Pennsylvania, Philadelphia, PA 19104-6396, USA  \\
         $^2$Complex Assemblies of Soft Matter, CNRS-Solvay-UPenn UMI 3254, Bristol, PA, 19007-3624, USA 
}

\date{\today}

\begin{abstract}
We introduce the concept of a ``hyperuniformity disorder length" $h$ that controls the variance of volume fraction fluctuations for randomly placed windows of fixed size.  In particular, fluctuations are determined by the average number of particles within a distance $h$ from the boundary of the window.  We first compute special expectations and bounds in $d$ dimensions, and then illustrate the range of behavior of $h$ versus window size $L$ by analyzing several different types of simulated two-dimensional pixel patterns -- where particle positions are stored as a binary digital image in which pixels have value zero/one if empty/contain a particle.  The first are random binomial patterns, where pixels are randomly flipped from zero to one with probability equal to area fraction.  These have long-ranged density fluctuations, and simulations confirm the exact result $h=L/2$.  Next we consider vacancy patterns, where a fraction $f$ of particles on a lattice are randomly removed.  These also display long-range density fluctuations, but with $h=(L/2)(f/d)$ for small $f$, and $h=L/2$ for $f\rightarrow 1$.  And finally, for a hyperuniform system with no long-range density fluctuations, we consider ``Einstein patterns'' where each particle is independently displaced from a lattice site by a Gaussian-distributed amount.  For these, at large $L$, $h$ approaches a constant equal to about half the root-mean-square displacement in each dimension.  Then we turn to grayscale pixel patterns that represent simulated arrangements of polydisperse particles, where the volume of a particle is encoded in the value of its central pixel.  And we discuss the continuum limit of point patterns, where pixel size vanishes.  In general, we thus propose to quantify particle configurations not just by the scaling of the density fluctuation spectrum but rather by the real-space spectrum of $h(L)$ versus $L$.  We call this approach ``Hyperuniformity Disorder Length Spectroscopy" (HUDLS). 
\end{abstract}

%
%
\maketitle

%
%
%
%

The structure of crystals is straightforward to describe.  By contrast, it remains a fundamental problem to identify and characterize the structural features of disordered many-body systems that relate to materials properties.  In this regard the concept of hyperuniformity was developed for systems in which long-wavelength density fluctuations are suppressed to nearly the same extent as in crystals \cite{TorquatoPRE2003, ZacharyJSM2009}.  Disordered systems that are hyperuniform accordingly possess some degree of long-ranged hidden order, not evident in the local structure, and this can give rise to special material properties.  For example, hyperuniformity has been connected to jamming \cite{DonevPRL2005, TorquatoStillingerRMP10, KuritaPRE2010, XuSM2010, BerthierPRL2011, ZacharyPRL2011, KuritaPRE2011, IkedaPRE2015, DreyfusPRE2015, WuPRE2015} and to novel photonic behavior \cite{FlorescuPNAS2009, ManPNAS2013, MullerAOM2014, ManOE16, Scheffold2016}  in amorphous materials.  It is relevant for the spatial arrangement of certain biological \cite{JiaoPRE2014} and astrophysical \cite{GabrielliPRD2002} objects.  And it has been connected with irreversibility in seemingly-deterministic many-body systems subjected to periodic driving \cite{HexnerPRL2015, TjhunPRL2015, WeijsPRL2015}, as well as to dynamical behavior of Brownian particles subjected to steady driving \cite{JackPRL2015}.  Thus the importance of quantifying hyperuniformity is increasingly recognized.

According to the original definition, a $d$-dimensional system of point particles is said to be hyperuniform if the structure factor $S(q)$ vanishes at wavevector $q=0$. \cite{TorquatoPRE2003}.  Equivalently, it is hyperuniform if the variance in the number of particles enclosed by a set of randomly-placed measuring windows of volume $\propto L^d$ grows as the average number of particles on the window surface, ${\sigma_N}^2(L)\sim L^{d-1}$ \cite{TorquatoPRE2003}.   Here $L$ is the width or diameter of the window and $d$ is dimensionality.  By contrast, the number variance for liquid-like systems with long-range density fluctuations grows like the average number particles inside the entire window, ${\sigma_N}^2(L)\sim L^{d}$; this corresponds to $S(0)>0$.  For extended particles of nonzero volume, the quantities of interest are instead the spectral density $\chi(q)$ and the variance ${\sigma_\phi}^2(L)$ of volume-fraction fluctuations \cite{ZacharyJSM2009, BerthierPRL2011, ZacharyPRL2011}.  The usual focus is on the small-$q$ / large-$L$ scaling;  i.e.\ on
\begin{align}
	\label{q_epsilon} \chi(q) & \sim q^{\epsilon}, \\
	\label{L_epsilon} {\sigma_\phi}^2(L) & \sim 1/L^{d+\epsilon},
\end{align}
and the value of the exponent.  For $\epsilon=0$, then there are long-ranged density fluctuations and the particles have a liquid-like arrangement that is significantly more random than in a crystal.  For $\epsilon>0$, the particles are arranged more uniformly throughout space and the system is said to be hyperuniform.  The degree of hidden order increases with $\epsilon$, and the upper limit $\epsilon=1$ corresponds to a crystal-like uniformity.  Logarithmic corrections to Eqs.~(\ref{q_epsilon}-\ref{L_epsilon}) may be expected, and can be described by a single value of $\epsilon$ over a (perhaps experimentally-limited) range of $q$ or $L$.

To relate to structure, it is helpful to translate measured quantities into a length scale.  For example Hopkins {\it et al.}~\cite{HopkinsPRE2012} define a correlation length from the value of $S(0)$.  And Refs.~\cite{HexnerPRL2015, WeijsPRL2015, WuPRE2015} define other lengths based on the location of features in the form of $\chi(q)$ versus $q$ or of $ {\sigma}^2(L)$ versus $L$.  Here, we introduce a completely length based on the value of ${\sigma_\phi}^2(L)$ that we call the ``hyperuniformity disorder length'', $h(L)$.  The size of $h(L)$ relates directly to the distance from the boundary over which fluctuations occur for a set of measuring windows, and its {\it value} -- as well as its scaling with $L$ -- indicates the degree of uniformity.

\begin{figure}[ht]
\includegraphics[width=2.0in]{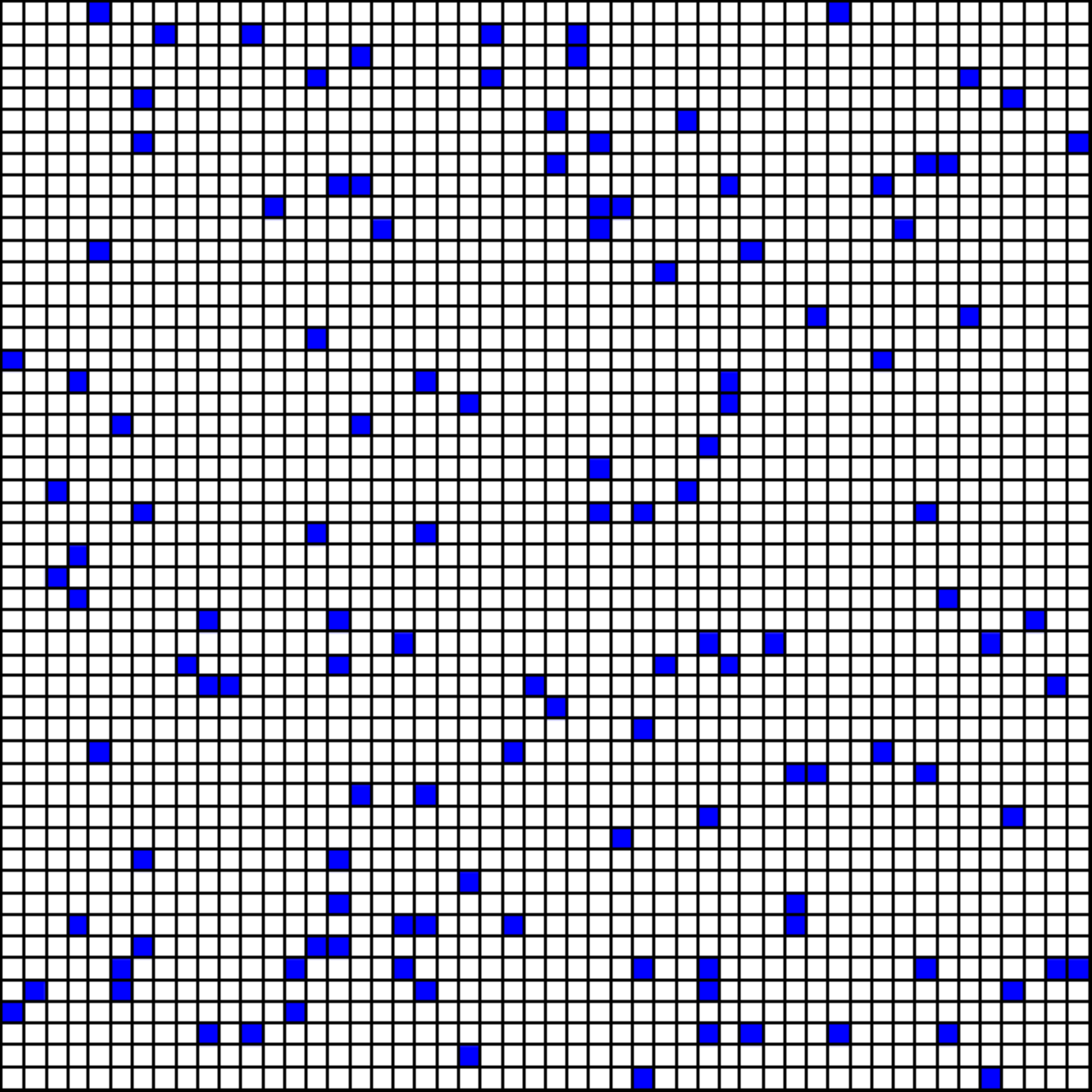}
\caption{(color online) Example binary pixel pattern, consisting of a $50\times 50$ grid of pixels and 125 particles (solid blue) occupying an area fraction of $\phi=125/(50\times50)=0.05$.}
\label{pixelpatterngrid}
\end{figure}

Particle positions, whether from experiment or simulation, can be represented by a digital image where pixel size is determined by experimental or numerical precision.  Accordingly, we develop our approach for $d$-dimensional pixel patterns.  A binary 2-dimensional example is shown in Fig.~\ref{pixelpatterngrid}, where space is represented by a square grid of pixels that have value $I(x,y)=0$ if empty or $I(x,y)=1$ if a particle is present.   We begin by deriving the volume fraction variance for a few special cases.  Using this as a guide, we discuss how to quantify hidden order from the measured variance, first by its ratio to that in a totally random arrangement of the same objects, and then by the hyperuniformity disorder length.  The continuum limit of vanishing pixel size, i.e. of point particles, is also discussed.  Then we illustrate this program for several classes of system, starting with binary pixel patterns and ending with grayscale pixel patterns representing a polydisperse collection of extended particles.


\section{Variance for Pixel Patterns}

In this section we define notation and compute the volume fraction variance exactly for four special arrangements of pixel particles (Poisson patterns, multinomial patterns, separated-particles, and cubic crytals).  As in Fig.~\ref{pixelpatterngrid}, the patterns to be characterized consist of a $d$-dimensional Cartesian grid of pixels with intensity values $I(x,y,z\ldots)$ that specify the arrangement of pixel-sized particles.  The ``volume" of a particle of species $i$ is considered to be $v_i = I_i {p_o}^d$, where $p_o$ is the pixel width and ${p_o}^d$ is the pixel volume.  Then the volume fraction $\phi$ equals the average intensity in the pattern.  For example, a system of extended particles with actual geometrical volumes $v_i$ can be represented by a pixel pattern where the central pixel inside each particle is incremented by intensity $I_i=v_i/{p_o}^d$.  In general, if species $i$ has number density $\rho_i$, then it occupies volume fraction $\phi_i = v_i \rho_i = I_i {p_o}^d \rho_i$ and the total volume fraction occupied by all particles is $\phi = \sum \phi_i$.  The average probability for a pixel to contain a particle of species $i$ is $q_i = {p_o}^d\rho_i = \phi_i/I_i$, to be used below.  

The basic tool for quantifying the extent of hidden order and hyperuniformity is the volume fraction variance ${\sigma_\phi}^2(L)$ versus window size.  Operationally, a large number of measuring windows of volume $V_\Omega$ are placed at random.  In each window $\Omega$, the volume fraction is computed as the average intensity. The results vary from window to window, but the average must converge to the total/true volume fraction $\phi$ if enough windows are sampled.  The variance of the measured volume fraction values, on the other hand, depends on the particular pattern and on the window size/shape.  The nature of the hidden order is to be analyzed from this behavior.

For analysis of simulated pixel patterns, we use cubic measuring windows of volume $V_\Omega=L^d$.  This will be useful for future analyses of digital video data.  For continuum space, it is more usual to use spherical measuring windows.  However this is inconvenient for pixel patterns, because spheres cannot be constructed except as pixelated approximates.  In this section, results for arbitrary window shapes are given in terms of $V_\Omega$ and results for hypercubic windows are given in terms of $L$.  Continuum results for hyperspherical windows of diameter $D$ are given in the conclusion.  The behavior of $h(L)$ for hypercubic windows is almost the same as $h(D)$ for hyperspherical windows.

\subsection{Random Patterns and Relative Variance}\label{randvar}

The volume fraction variance may be computed exactly, as follows, if the pixel particles are arranged at random.  For a given $L^d$ measuring window of $n=(L/p_o)^d$ pixels, the total volume of pixel particles is $V = \sum v_i N_i$ where $N_i$ is the actual number of species $i$ enclosed (i.e. the number of time enclosed pixels were incremented by $+I_i$).  The corresponding average and variance over window locations are $\overline V=\sum v_i \overline{N}_i$ and ${\sigma_V}^2 = \sum v_i^2 {\sigma_{N_i}}^2 + \sum_{i\ne j} v_iv_j\sigma_{N_iN_j}$, where $\overline{N}_i=\rho_i L^d$ is the average number of species $i$ enclosed, ${\sigma_{N_i}}^2$ is the variance, and $\sigma_{N_iN_j}=\overline{N_iN_j}-\overline{N}_i\overline{N}_j$ is the covariance.

To evaluate the variance and covariance for the number of enclosed particles, we must now make a distinction between two different kinds of random configuration.  If multiple particles can occupy the same pixel, then Poisson statistics hold, the number variance is ${\sigma_{N_i}}^2=n q_i$ where the probability $q_i$ was given above, and the covariance is zero.  In this case, each particle's location is chosen at random and multiple particles may freely occupy a pixel.  Such configurations are called ``Poisson" patterns.  By contrast, if only one particle at a time is allowed on each pixel, then multinomial statistics hold: the number variance is ${\sigma_{N_i}}^2=n q_i(1-q_i)$ and the covariance is $\sigma_{N_iN_j}=-nq_iq_j$.  This case corresponds to a pattern where the intensity of each pixel is randomly drawn from $\{0, I_1, I_2,\ldots\}$ with respective probabilities $\{1-\sum q_i, q_1, q_2,\ldots\}$.  We call such configurations ``multinomial" patterns.

Combining these ingredients, the volume fraction variance for the two different types of random pattern is computed to be
\begin{numcases}
   { {\sigma_\phi}^2(L)  = \frac{\langle I\rangle {p_o}^d}{L^d}   }
	\phi & {\rm Poisson,} \label{spoiss} \\
	\phi\left[1-{\phi}/{\langle I \rangle}\right] & {\rm Multinomial,} \label{smulti}
\end{numcases}
where $\langle I \rangle = \sum \phi_i I_i/\phi$ is the volume-fraction weighted average grayscale intensity of the particle species.  The common prefactor may be re-written as $\langle v\rangle/V_\Omega$ where $\langle v\rangle = \langle I\rangle {p_o}^d$ is the volume-fraction weighted particle volume and $V_\Omega$ is the window volume -- no matter what its shape.  Thus, the variance scales as $1/L^d$ as expected for a system with long-range density fluctuations.  And, importantly for later, the actual proportionality constants are now known.  In principle, Eqs.~(\ref{spoiss}-\ref{smulti}) could be obtained from reciprocal-space results~\cite{LuTorquatoJCP90, QuintanillaTorquatoJCP97, QuintanillaTorquatoJCP99, ZacharyJSM2009}.

To our knowledge, neither pixelated space nor the distinction between Poisson and multinomial statistics have been previously considered.  This is important when a large fraction of pixels contain particles.  For example, a random ``binomial pattern" of non-overlapping binary particles, with $I=1$, has variance ${\sigma_\phi}^2(L)=(\langle v\rangle/V_\Omega) [\phi(1-\phi)]$; this holds at all $\phi$, and is correctly unchanged by image inversion $I\rightarrow 1-I$ and $\phi\rightarrow 1-\phi$.  Note that Poisson statistics are recovered to good approximation when the populated pixels are dilute and the $q_i$ are small compared to one.  This happens for a central-pixel representation when the image resolution is good, i.e.\ in the continuum limit where the pixel size is small enough that $\phi \ll \langle v\rangle/{p_o}^d = \langle I \rangle$ is true.

On the basis of Eqs.~(\ref{spoiss},\ref{smulti}) we define the {\it relative variance} for general configurations as
\begin{numcases}
   { \mathcal V(L) \equiv  }
	\frac{ {\sigma_\phi}^2(L) }{\phi} & {\rm Poisson,} \label{vpoiss} \\
	\frac{ {\sigma_\phi}^2(L) }{\phi[1-\phi/\langle I\rangle]}  & {\rm Multinomial,} \label{vmulti}
\end{numcases}
where the two cases are respectively for particles that are, or are not, allowed to overlap.  Then for both types of random pixel patterns, the relative variance is $\mathcal V(L) = \langle v\rangle/V_\Omega$ for any volume fraction.  This equals $\langle I\rangle$ at the smallest window size, $V_\Omega={p_o}^d$, and has an initial linear decay of $\mathcal V(L) = \langle I\rangle[ 1 - d(L-p_o)/p_o + \mathcal O(L-p_o)^2]$ for hypercubic windows.  For patterns with hidden order, $\mathcal V(L)$ will be smaller than this upper bound \cite{bounds}.

\subsection{Separated-Particles Limit}

The volume fraction variance may also be computed exactly for small $L$, when all measuring windows contain no more than one particle.  For example if the pixel particles effectively repel each other, or if for another reason they have some minimum separation, then for a set of $w$ randomly placed $L^d$ measuring windows that are sufficiently small, the results will be some numbers $\{M_o, M_1, M_2,\ldots\}$ of observed intensity values $\{0, I_1, I_2,\ldots\}$.  Then the measured volume fraction moments are $\phi_m^n =\sum (M_i/M)(I_i {p_o}^d/L^d)^n$, where $M=\sum M_i$,  and the measured relative variance is
\begin{equation}
	\mathcal V_m(L) = \frac{  \sum  \frac{M_i}{M\phi} \left( I_i {p_o}^d /L^d \right)^2 - \phi} {1-\phi/\langle I\rangle },
\label{VmSmallL}
\end{equation}
assuming multinomial statistics as appropriate for separated particles.  The averages are set by $\overline M_i = M\rho_i L^d = M[\phi_i/(I_i {p_o}^d)]L^d$.  This gives $\overline {\phi_m^n} = \sum \phi_i( I_i {p_o}^d/L^d)^{n-1}$, which leads to $\overline {\phi_m} = \sum \phi_i = \phi$ and a relative volume fraction variance of
\begin{eqnarray}
	\mathcal V(L) &=& \langle I\rangle \frac{ (p_o/L)^d - \phi/\langle I \rangle}{1-\phi/\langle I\rangle}, \label{VsmallL}
\end{eqnarray}
Note that Eq.~(\ref{VsmallL}) is exact and holds for small enough $L$, for {\it any} arrangement of particles.  If the arrangement is random, then it holds only for $L=p_o$ since there will be a nonzero number of adjacent pixel particles.  If the particles are separated, then Eq.~(\ref{VsmallL}) holds up to some larger $L$ that is set by the smallest particle-particle separation.  For windows of arbitrary shape, Eq.~(\ref{VsmallL}) generalizes to $\mathcal V(L) = (\langle v\rangle/V_\Omega-\phi)/(1-\phi/\langle I\rangle)$.  To our knowledge, the small window limit of separated particles has not been previously considered.

\subsection{Cubic Crystal}

As a third example that may be computed exactly, we consider a pattern consisting of pixel particles of volume $v=I_o{p_o}^d$ on a $d$-dimensional cubic lattice of spacing $b$, using $L^d$ hypercubic measuring windows, where both the lattice and the windows are aligned with the grid of pixels.  For this situation there are only $(b/p_o)^d$ distinct locations for the windows; and there are only $d+1$ distinct results for the number $n$ of enclosed particles.  These are all of form $n=\sum_{j=0}^d a_j [{\rm floor}(L/b)]^j$, with $a_d=1$.  For example, in $d=1$ there are $(b-\delta)/p_o$ ways for $n_1={\rm floor}(L/b)$ particles to be enclosed, and $\delta/p_o$ ways for $n_1+1$ particles to be enclosed, where $\delta=L-b{\rm floor}(L/b)$.  The probability distribution and the moments for the number of enclosed particles may then be evaluated.  Exact results for the average and variance of the measured volume fractions are found by direct summation to be $\phi=I_o(p_o/b)^d$, as expected, and
\begin{widetext}
\begin{equation}
  {\sigma_\phi}^2(L) = \phi^2\left( \frac{b}{L} \right)^{2d}
      	\left\{ \frac{L}{b}-{\rm floor}\left(\frac{L}{b}\right)\left[ 1 - 2 \frac{L}{b}+ {\rm floor}\left(\frac{L}{b}\right)\right] \right\}^d - \phi^2~.
\label{cubes}
\end{equation}
\end{widetext}
We calculated Eq.~(\ref{cubes}) explicitly in 1, 2, and 3 dimensions, and we verified the $d=2$ case by comparison with simulation.  Though we are unaware of prior statement or proof, we suppose it is true in all dimensions.  The corresponding number variance is ${\sigma_N}^2(L)=[{\sigma_\phi}^2(L)/\phi^2](L/b)^{2d}$.  For $L<b$ the results simplify to ${\sigma_\phi}^2(L)=\phi^2[(b/L)^d-1]$ and ${\sigma_N}^2(L) = (L/b)^d[1-(L/b)^d]$.  The former agrees with Eq.~(\ref{VsmallL}); the latter agrees with the $d=1$ case considered in Ref.~\cite{TorquatoPRE2003}, where the number variance was computed from the structure factor and expressed in Eq.~(83) as a Fourier series.  For increasing $L$, the variance vanishes at integer values of $L/b$.  The variance at half-way between the zeros gives the large-$L$ decay envelope as ${\sigma_\phi}^2(L)=(d/4)(\phi b/L)^2$.  This is pathological for $d>1$, since crystals ought to be strongly hyperuniform with variance scaling of $1/L^{d+1}$.  Our result is even more pathological than expected based on footnote~11 on p.~14 of Ref.~\cite{GabrielliPRD2002}, which indicates that ${\sigma_\phi}^2(L)\sim1/L^{d-1}$ was found for cubic crystals with cubic measuring windows.  The pathology arises because the distribution of measured $\phi_{\Omega}$ values is highly non-Gaussian; hence, it may be removed by using spherical measuring windows \cite{TorquatoPRE2003, GabrielliPRD2002}.  For pixelated space it is more convenient to tilt the lattice at multiple angles with respect to the pixel grid, as done in the simulations below.

\subsection{Continuum Limits}

The above expectations were all derived for pixelated space, where window widths and lattice spacings are an integer number of pixel widths $p_o$, and where each particle has intensity $I_i$ and occupies a ${p_o}^d$ voxel.  But the results all also extend to continuous space for a point representation of particles of volume fraction $\phi$ and $\phi_i$-weighted average particle volume $\langle v\rangle.$  This is the limit of vanishing $p_o$ and diverging $I_i$ taken simultaneously such that $v_i=I_i{p_o}^d$ and $\phi_i$ are constant. Then multinomial statistics reduce to Poisson statistics, since $\phi/\langle I\rangle$ vanishes and there is zero probability for two point particles to lie on top of one another.  For measuring windows of any shape, and volume $V_\Omega$, the relative variance expectation of Eq.~(\ref{spoiss}) for a random arrangement then becomes ${\sigma_\phi}^2(L)/\phi=\langle v\rangle/V_\Omega$.  For an infinite system, where there is no minimum particle-particle separation, this holds for all $V_\Omega$.  For a finite system, however, it fails for small $V_\Omega$ where the windows all contain no more than one particle.  For {\it any} pattern and window shape at such small $V_\Omega$, the relative variance expectation from Eq.~(\ref{VsmallL}) is ${\sigma_\phi}^2(L)/\phi=\langle v\rangle/V_\Omega - \phi$.  For cubic crystals, Eq.~(\ref{cubes}) holds in the continuum limit as written.


\section{Characterizing Hidden Order}

We now discuss how to measure the volume fraction variance for a given pixel pattern, then we propose two different ways to plot and interpret the results.

\subsection{Measurements and Errors}

Standard procedure is to measure the variance of the list of volume fractions found inside a large number $w$ of randomly placed measuring windows $\Omega$ of desired size and shape \cite{TorquatoPRE2003}.  For pixel patterns, the volume fraction inside a particular measuring window is $\phi_m = [\sum I(x,y,z...){p_o}^d]/V_\Omega$, which equals the total enclosed particle volume divided by the measuring window volume $V_\Omega\sim L^d$; this is the same as the average intensity of all the enclosed pixels.  The average $\overline{\phi_m}$ and the variance ${\sigma_m}^2=\overline{\phi_m^2}-\phi^2$ of the measured volume fractions may then be computed.  For a large number $w$ of windows, $\overline{\phi_m}$ will converge to the actual total volume fraction $\phi$ of the whole pattern.  But how large must $w$ be, and what is the resulting expected statistical uncertainty between the measured variance and the true value?  To our knowledge, this has not been addressed in prior work.

The error analysis for large measuring windows is most straightforward.  Since then the distribution of measured volume fractions ought to be Gaussian, the uncertainty can be estimated from the standard error of the variance as $\Delta {\sigma_\phi}^2(L) = {\sigma_\phi}^2(L) \sqrt{2/(s-1)}$, where $s$ is the number of independent samples.  If the $w$ measuring windows are small and the pattern is large, then the windows measure disjoint sets of particles and hence $s=w$ may be used.  But for measuring windows that are sufficiently large in size or number, some of the particles will be sampled by more than one window and hence $s$ will become smaller than $w$.  In general, the number $s$ of independent samples will be the total number of pixels covered by the entire set of measuring windows divided by the number of pixels per window.  To account for this we create an auxiliary binary image where pixel values $I_{aux}(x,y)$ are flipped from zero to one if the corresponding pixel in the image lies within a measuring window.  Then the number of independent samples is taken from the auxiliary image as $s=(\sum I_{aux})/(L/{p_o})^d$.

To cover most of the sample with windows, there is no need for $w$ to be more than on the order of $V/V_\Omega$ where $V$ is the volume of the entire image.  This gives a number of independent samples that is close to $s_{max}=V/V_\Omega$.  As window size increases toward system size, $s_{max}$ decreases to one and causes the standard error of the variance to bloom.  In our analyses of two-dimensional simulations, below, we take the number of $L\times L$ square windows to be $w=A/(2L^2)$, where $A$ is the image area, subject to the constraint $10^2\le w \le 10^4$.  And the largest window size we use is typically $L=\sqrt{A}/2$.  Results for larger measurement windows are contaminated by finite size artifacts, such that the area-fraction variance is artificially depressed toward zero as $L$ approaches the image width \cite{DreyfusPRE2015}.  Care must be taken that this systematic error not be mistakenly interpreted as a sign of hyperuniformity or lack thereof.
 
For small measuring windows, the standard error of the variance is not a good estimate because the volume fraction distribution is not Gaussian.  Instead, the statistical uncertainty may be deduced from Eq.~(\ref{VmSmallL}), which gives $(\Delta \mathcal V)^2 \propto \sum (\Delta w_i)^2$ where $w_i$ is the number of small measuring windows that contain one particle of species $i$.  The key ingredient is $(\Delta w_i)^2 = \overline{w_i} = w\rho_i V_\Omega$.  As before, $w$ should be replaced with $s$.  The resulting formula for the uncertainty is good only for small enough windows that contain no more than one particle.  But we suppose it might serve as an estimate for intermediate window size, anyway.  Combining this in quadrature with the standard error of the variance, which applies for large windows, gives the expected statistical uncertainty in the measured relative variance as
\begin{equation}
	\Delta \mathcal V =  \sqrt{  \frac{ \langle v^3\rangle /{V_\Omega}^{3} }{ s\phi[1 -\phi/ \langle I\rangle ]^2 }  + \frac{2\mathcal V^2}{s-1}  }~,
\label{error}
\end{equation}
assuming multinomial statistics.  For Poisson statistics, the continuum limit, or pixel particles with $\phi\ll\langle I\rangle$, the $-\phi/\langle I\rangle$ term is dropped in comparison with one.  Note that the first term dominates for small windows, while the second dominates for large windows; therefore, it does no harm to simply add them together.  We will use Eq.~(\ref{error}) to generate error bars in analysis plots, below.  Note that the $\phi_i$-weighted moments of the particle volume distribution may be written as $\langle v^n\rangle = \sum \phi_i {v_i}^n/\phi = (\sum N_i {v_i}^{n+1}) / (\sum N_i {v_i})$ where the sums are over species and where $N_i$ is the total number of particles of species $i$ in the sample.  This may also be written as $\langle v^n\rangle = \sum {v_j}^{n+1} / \sum {v_j}$ where the sums are over all particles in the sample.

\subsection{Variance Ratio, $\mathcal R(L)$}

The volume fraction variance is largest for a totally random pattern, and necessarily decreases with ordering.  Therefore the presence of hidden order can be detected by comparing ${\sigma_\phi}^2(L)$ for the pattern in question with $[{\sigma_\phi}^2(L)]_{rand}$ for a totally random arrangement of the {\it same} set of particles.  For non-overlapping pixel particles, the variance at $L=p_o$ merely measures the distribution of particle intensities independent of their arrangement; therefore, ${\sigma_\phi}^2(p_o) = [{\sigma_\phi}^2(p_o)]_{rand} = \phi\langle I\rangle[1-\phi/\langle I\rangle]$ holds and we may write $[{\sigma_\phi}^2(L)]_{rand} = {\sigma_\phi}^2(p_o)(p_o/L)^d$.  Hence we define the variance ratio, and evaluate it without having to know $\phi$ and $\langle I\rangle$, as
\begin{equation}
	\mathcal R(L) \equiv     \frac{{\sigma_\phi}^2(L) }{[{\sigma_\phi}^2(L)]_{rand}} =  \frac{ {\sigma_\phi}^2(L) L^d}{{\sigma_\phi}^2(p_o) {p_o}^d}.
\label{rrat}
\end{equation}
By construction, totally random patterns have $\mathcal R(L)=1$ and this is an upper bound.  Note, too, that the variance ratio is normalized to $\mathcal R(p_o)=1$ for all pixel patterns.   For the continuum limit of point particle and spherical measuring windows, simply replace $L$ with diameter $D$, and $p_o$ with some chosen smallest measuring diameter $D_o$ that is less than the minimum particle-particle separation $r_{min}$ in the sample.  Either $D_o$ should be vanishingly small compared to $r_{min}$, or else a correction should be made using the above results for separated-particle arrangements:
\begin{equation}
	\mathcal R(D) = \frac{ {\sigma_\phi}^2(D)D^d }{  [ {\sigma_\phi}^2(D_o)+\phi^2]{D_o}^d  }.
\label{rcontinuum}
\end{equation}
This gives $\mathcal R(D)=1-\phi V_\Omega/\langle v\rangle$ exactly for $D<s$, and is normalized to $\mathcal R(D_o)=1$ only as $D_o$ vanishes and ${\sigma_\phi}^2(D_o)$ diverges.  To interpret $\mathcal R(L)$, first note that the large-$L$ scaling is $\mathcal R(L)\sim 1/L^\epsilon$ for ${\sigma_\phi}^2(L) \sim 1/L^{d+\epsilon}$.  Therefore, $\mathcal R(L)$ goes to a constant for patterns with long-range density fluctuations ($\epsilon=0$); and it goes to zero as a dimension-indepenent power-law if the pattern is hyperuniform, e.g. most notably $\mathcal R(L)\sim 1/L$ for strongly hyperuniform ($\epsilon=1$).  The utility of diagnosing hyperuniformity via the product ${\sigma_\phi}^2(L)L^d$ has been previously recognized \cite{DreyfusPRE2015, WuPRE2015}.  The advantage of additionally normalizing by ${\sigma_\phi}^2(p_o){p_o}^d$ is that then the {\it value} (not just the scaling) has meaning.  Smaller $\mathcal R$ means more order, larger $\mathcal R$ means more random, and $\mathcal R=1$ means totally random.   Thus, $\mathcal R(L)$ can be thought of as a randomness index, which decays from one as $L$ increases and hidden order is detected.  As shown below, $\mathcal R(L)$ can also be interpreted as the fraction $f$ of space available for density fluctuations at wavelength $L$.

\subsection{Hyperuniformity Disorder Length, $h(L)$}

\begin{figure}[ht]
\includegraphics[width=2.000in]{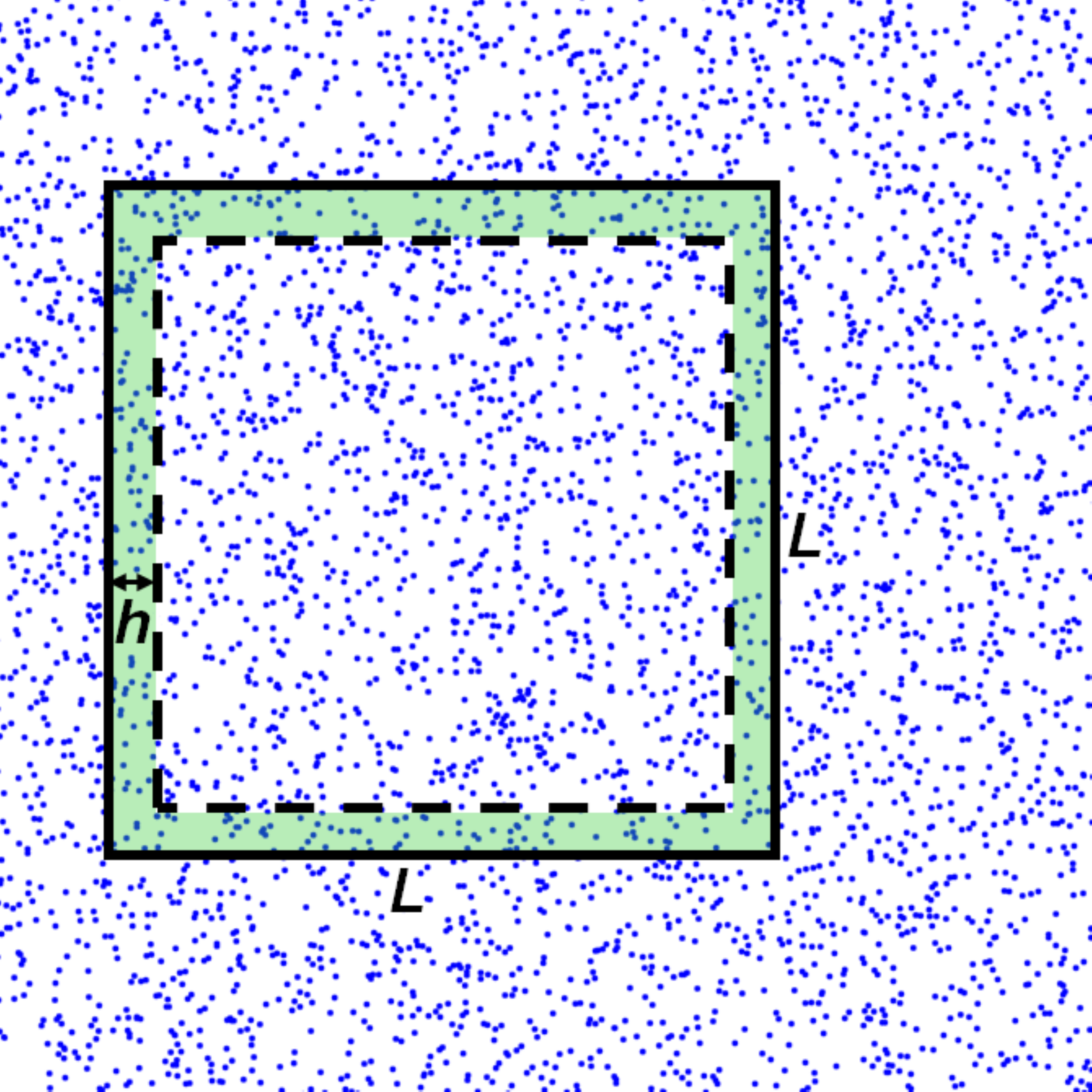}
\caption{(color online) Example ``Einstein pattern" of pixel particles with a Gaussian-distributed random displacement from a triangular lattice.  The lattice constant is 30 pixels, and the root-mean-square displacement is 150 pixels in each dimension.  Since the particles are effectively bound to the lattice sites, the pattern is hyperuniform with no long-range density fluctuations -- even though it appears very disordered to the eye.  Randomly placed measurement windows hence show density fluctuations that are due only to particles within a distance $h$ of the window boundary that is constant for large windows.  Here this ``hyperuniformity disorder length'' is $h=82$ pixels, as illustrated for an example $1100\times1100$ window.  For clarity, the pixel particles are shown as dots that are 11 pixels across.}
\label{windowschematic}
\end{figure}

While $\mathcal R(L)$ is a useful new quantity, it does not connect to the original idea \cite{TorquatoPRE2003} that fluctuations in hyperuniform systems are governed by particles on the {\it surface} of the measuring windows.  Since particles reside in a volume, we quantify this notion by introducing a hyperuniformity disorder length $h(L)$ that specifies the region near the window boundary where fluctuations are important.  This concept is depicted in Fig.~\ref{windowschematic}, which shows an $L\times L$ measuring window that is partitioned into a boundary region of thickness $h$ and an $(L-2h)\times(L-2h)$ interior region.  Intuitively, $h$ is defined such that, if the system is ergodic and the time variation $\phi_{\Omega}(t)$ for one window has the same distribution as for a randomly placed set of windows, then only the boundary particles have opportunity to temporarily leave the measuring window.  For a totally random system where all enclosed particles participate, $h(L)$ would equal the maximum value $L/2$.  For a ``maximally" hyperuniform system, $h(L)$ would be a minimum and equal to $p_o/2$, which is the largest constant consistent with the $h\le L/2$ requirement for all $L$; in this case, the participating pixel particles are literally on the surface.  Since $h(L)$ increases with disorder, we call it the hyperuniformity disorder length.  In general, the nature of the fluctuations is thus easily visualized in terms of the value and form of the real-space spectrum of $h(L)$ versus $L$.  We therefore call this analysis method ``Hyperuniformity Disorder Length Spectroscopy" (HUDLS).  

The technical definition of $h(L)$ is based on the measured variance compared to that for a totally random arrangement of the same set of particles.  For a truly random pattern, the number of pixels with fluctuating particles is equal to the number $n=(L/p_o)^d$ of pixels in the entire $L^d$ measuring window.  For non-random patterns with smaller variance, we take it instead to be the number $n_B=[L^d-(L-2h)^d]/{p_o}^d$ of pixels in the boundary region; this is where $h$ enters.  Intuitively, the particle number variance is set by the average number of particles in the boundary region.  Repeating the arguments in the first three paragraph of Section~\ref{randvar}, but with $n$ replaced by $n_B$ as the only difference, gives the following {\bf Fundamental Equations of HUDLS} for cubic windows:
\begin{eqnarray}
	\mathcal V(L) &=& \frac{\langle v\rangle}{L^d} \left[ \frac{L^d - (L-2h)^d}{L^d} \right], \label{vgen} \\
	\mathcal R(L) &=& 1 -\left( 1 - {2h}/{L}\right)^d, \label{rgen} \\
	h(L) &=& (L/2)\{ 1 - \left[  1 - \mathcal R(L) \right]^{1/d}\}. \label{hgen}
\end{eqnarray}
Note that the term in square brackets in Eq.~(\ref{vgen}) equals the right-hand side of Eq.~(\ref{rgen}), and is the ratio of boundary volume to window volume.  As defined earlier, $\mathcal V(L)$ is shorthand for the relative volume fraction variance given by Eqs.~(\ref{vpoiss}-\ref{vmulti}) for either Poisson or multinomial statistics, as appropriate;  $\langle v\rangle = \langle I\rangle {p_o}^d$ is the $\phi_i$-weighted average particle volume; and $\mathcal R(L) \equiv \mathcal V(L)/[ \langle v\rangle/L^d]$ is the ratio of the variance to that in a random arrangement of the same particles.  We emphasize that Eqs.~(\ref{vgen}-\ref{hgen}) are equivalent, and serve to define $h(L)$ in terms of the measured volume fraction variance.  The same set of equations holds for point particles in continuum space, and also for spherical measuring windows if $L$ is replaced by diameter.


Before putting this machinery into action, we examine special cases and bounds.  First, note that $\mathcal V(L)=\langle v\rangle/L^d$ and $\mathcal R(L)=1$ are recovered for totally random patterns, where $h$ equals $L/2$ by construction.  These are upper bounds \cite{bounds}.  For not-totally-random patterns with liquid-like long-range fluctuations, $\mathcal V(L)$ also scales as $1/L^d$; then $\mathcal R(L)$ goes to a constant and $h(L)$ scales like $L$ but with a proportionality constant less than 1/2.  For separated particles at small $L$, where there is no more than than one particle per window, the expectations are
\vspace{0.2in}
\begin{eqnarray}
	\mathcal V(L) &=& \frac{ \langle v\rangle/L^d - \phi}{1-\phi/\langle I\rangle} , \label{vsep} \\
	\mathcal R(L) &=& \frac{ 1 - \phi L^d/\langle v\rangle }{1-\phi/\langle I\rangle }, \label{rsep} \\
	h(L) &=& \frac{L}{2}-\frac{L}{2}\left[  \frac{ \phi L^d/\langle v\rangle - \phi/\langle I\rangle}{1-\phi/\langle I\rangle} \right]^{1/d}~. \label{hsep}
\end{eqnarray}
These are lower bounds; however, at larger $L$ where Eq.~(\ref{hsep}) falls below $p_o/2$, the lower bounds are given instead by the  fundamental equations evaluated at $h = p_o/2$ (maximally hyperuniform).  In the continuum limit, the $\phi/\langle I\rangle$ terms vanish.  For patterns where $h(L)\ll L$ holds and fluctuations are hence only near the measuring window surfaces, then expansion of the fundamental equations gives
\begin{eqnarray}
	\mathcal V(L) &=& 2d\frac{\langle v\rangle h(L)}{L^{d+1}},  \label{vsmallh} \\
	\mathcal R(L) &=& 2d \frac{h(L)}{L}, \label{rsmallh} \\
		        h(L) &=& \frac{1}{2d}\mathcal R(L)L. \label{hsmallh}
\end{eqnarray}
The case of strong hyperuniformity is $\mathcal V(L)\sim \langle v\rangle/L^{d+1}$; this corresponds to a constant hyperuniformity disorder length $h(L)=h_e$.  Note that the very form of the $\mathcal V(L)\sim \langle v\rangle/L^{d+1}$ scaling demands the existence of a new length scale, $h_e$, in order to become dimensionally correct: $\mathcal V(L) \propto \langle v\rangle h_e/L^{d+1}$.


For further intuition and direct connection with the surface coefficient $\Lambda$ of Ref.~\cite{TorquatoPRE2003}, we repeat the derivation of Eq.~(\ref{vsmallh}) for the restricted case of monodisperse particles of volume $v$, Poisson statistics for boundary fluctuations, arbitrary window shapes, and $h\ll {V_\Omega}^{1/d}$.  Then the boundary volume equals window surface area $A_\Omega$ times $h$, and the number variance equals the average number of particles in $A_\Omega h$:
\begin{equation}
	{\sigma_N}^2 = \frac{A_\Omega h \phi}{v}.
\label{sigmaN}
\end{equation}
The volume fraction variance is therefore
\begin{equation}
	{\sigma_\phi}^2 \equiv \left( \frac{v}{V_\Omega} \right)^2 {\sigma_N}^2 = \frac{v A_\Omega h \phi}{ {V_\Omega}^2 },
\label{sigmaV}
\end{equation}
and the relative variance is
\begin{equation}
	\mathcal V(L) \equiv \frac{ {\sigma_\phi}^2 }{\phi} = \frac{v A_\Omega h}{ {V_\Omega}^2 }.
\end{equation}
This reduces to Eq.~(\ref{vsmallh}) for hypercubic windows, where $V_\Omega = L^d$ and $A_\Omega = 2d L^{d-1}$.  In Ref.~\cite{TorquatoPRE2003} a surface coefficient is defined by ${\sigma_N}^2 = \Lambda(R/b)^{d-1}$ and tabulated for hyperspherical windows of radius $R$ and monodisperse crystals with lattice spacing $b$.  Thus $\Lambda$ and $h_e$ are directly related by $\Lambda(R/b)^{d-1} = A_\Omega h_e \phi/v$.


As another aside, it might be tempting to use the fundamental equations to analyze actual images of extended particles -- rather than their central-pixel representation.  We warn that this is correct only for sufficiently large windows.  Ref.~\cite{DJDhudls} generalizes our HUDLS analysis approach to extended particles, and corrects upon Eq.~(\ref{vgen}) at small $L$ by accounting for particles that lie partially inside and partially outside the measuring windows.

\subsection{Multinomial Expectations in $d=2$}

In the following sections we measure the area fraction variance for simulated two-dimensional patterns of non-overlapping pixel particles.  So multinomial statistics are appropriate, and for reference the three quantities of interest are
\begin{eqnarray}
	\mathcal V(L) &=& \frac{ {\sigma_\phi}^2(L) }{ \phi[1-\phi/\langle I\rangle] } = \frac{ 4 \langle a\rangle (L-h)h }{L^4}, \label{v2d} \\
	\mathcal R(L) &=& \frac{ {\sigma_\phi}^2(L) L^2 }{ {\sigma_\phi}^2(p_o){p_o}^2 }, \label{r2d} \\
	h(L) &=& (L/2)[1-\sqrt{ 1 - \mathcal R(L)] },    \label{h2d}
\end{eqnarray}
where $\langle a\rangle = \langle I\rangle {p_o}^2$ is the area-fraction weighted average particle area.  These equations will be used to deduce $\mathcal R(L)$ and $h(L)$ from measurements of $\mathcal V(L)$. Results will be compared with the following bounds:
\begin{widetext}
\begin{eqnarray}
	{\rm max}\left\{ \frac{\langle a\rangle/L^2-\phi}{1-\phi/\langle I\rangle},~  \frac{\langle I\rangle (2L/p_o-1)}{(L/p_o)^4}\right\}    &\le& \mathcal V(L) \le    \frac{\langle a\rangle}{L^2}, \label{vbounds} \\
	{\rm max}\left\{ \frac{1-\phi L^2/\langle a\rangle}{1-\phi/\langle I\rangle},~ \frac{\langle I\rangle (2L/p_o-1)}{(L/p_o)^2}\right\}     &\le& \mathcal R(L) \le   1, \label{rbounds} \\
	{\rm max}\left\{ \frac{L}{2p_o}\left(1-\sqrt{  \frac{\phi L^2/\langle a\rangle-\phi/\langle I\rangle}{1-\phi/\langle I\rangle}    }~\right),~  \frac{1}{2} \right\}   &\le& \frac{h(L)}{p_o} \le    \frac{L}{2p_o}. \label{hbounds}
\end{eqnarray}
\end{widetext}
The upper bounds are for totally random patterns; these will be plotted as red dashed lines.  The lower bounds are either the small-$L$ expectation for separated particles, or the expectation for maximally-hyperuniform as defined by $h=p_o/2$; these will be plotted as green dotted curves.  Note that, with substitution of $\langle a\rangle = \langle I\rangle {p_o}^2$, the bounds are all algebraic functions of the dimensionless variable $x=L/p_o$.  For small $L$, and for very disordered patterns, it can happen that the measured variance exceeds the upper bound in accord with expected statistical uncertainty.  Then $\mathcal R$ is greater than one and $h$ cannot be deduced from Eq.~(\ref{h2d}); instead, we take $h= (L/2)[1+\sqrt{\mathcal R -1}]$ as a reasonable way to show data points and error bars that overlap the upper bound.


\section{Binary Pixel Particles}

We now apply the above methods to characterize three different types of two-dimensional binary pixel patterns created by simulation.  Here the image sizes are all $8600p_o \times 8600p_o$, somewhat larger than typical digital video images.  Pixel values are either $I(x,y)=0$ (empty) or $I(x,y)=1$ (one particle); therefore multinomial statistics reduce to binomial statistics, all particle areas are $a={p_o}^2$, and the area fraction equals the fraction of pixels that contain a particle.  The area fraction variance ${\sigma_\phi}(L)^2$ is measured as described above using $L\times L$ measuring windows, where $L$ is varied from $p_o$ to at least $4300p_o$, half the image width.

\subsection{Random Binomial Patterns}

A simple algorithm is to flip each pixel from zero to one with probability equal to the desired area fraction, $\phi$.  This creates a random ``binomial'' pixel pattern, with an actual area fraction that is approximately $\phi$.  To create a binomial pattern with exactly the desired area fraction, we instead select a pixel at random, set its value to one, and repeat until the desired number of pixel particles is reached.  Such binomial patterns are the pixelated analogue of random Poisson patterns for point particles.  While points literally occupy zero volume in Poisson patterns, binomial patterns have $0<\phi<1$ and reduce to a Poisson pattern at both extremes.

\begin{figure}[ht]
\includegraphics[width=2.75in]{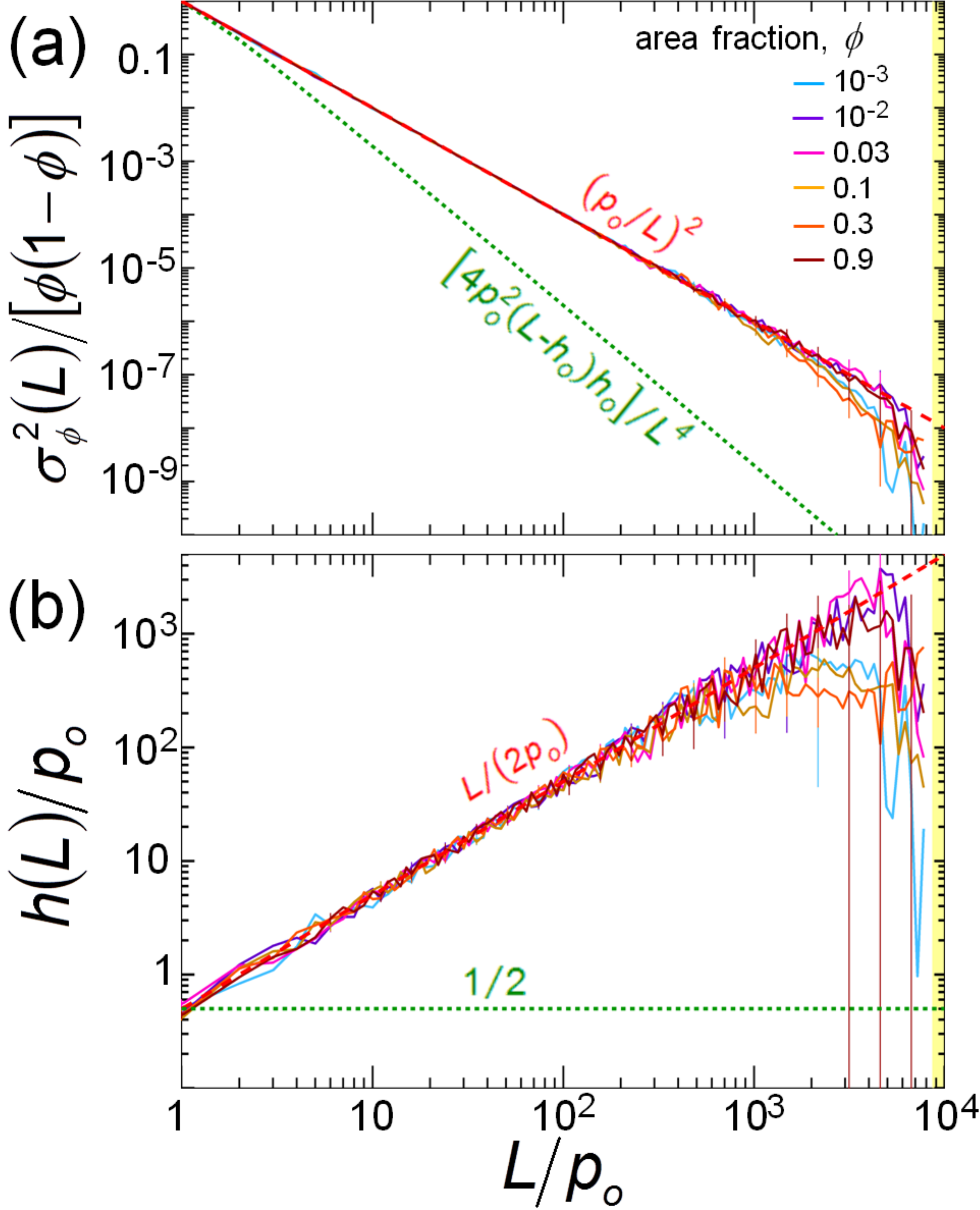}
\caption{(Color online) Area fraction variance (a) and hyperuniformity disorder length (b) versus measuring window size, for simulated two-dimensional random binomial pixel patterns with total area fraction $\phi$ as labeled.  The system size is $8600p_o \times 8600p_o$ where $p_o$ is the pixel width; this is indicated by yellow shading.  The results agree well with expectation of (a) ${\sigma_\phi}^2/[\phi(1-\phi)]=(p_o/L)^2$ and (b) $h=L/2$.  However there are significant finite-size effects for large $L$, so all further simulated patterns will be analyzed only up to half the system size.}
\label{binomial}
\end{figure}

Simulation results for the relative variance $\mathcal V(L)={\sigma_\phi}^2(L)/[\phi(1-\phi)]$ are plotted versus $L/p_o$ in Fig.~\ref{binomial}a for six different target area fractions, ranging widely from $\phi=10^{-3}$ up to $\phi=0.9$.  All data collapse together and match the expectation $(p_o/L)^2$ to within statistical uncertainty for $L$ less than about half the image width, where finite-size effects are expected to become strong \cite{DreyfusPRE2015}.  The corresponding hyperuniformity disorder lengths, deduced from Eq.~(\ref{h2d}), are plotted underneath in Fig.~\ref{binomial}b.  All results collapse together and match the expectation $h=L/2$ to within statistical uncertainty, as long as finite-size effects are absent.  The good agreement between simulation and expectation validates our calculations as well as our image analysis procedures.  Note in particular that the factor of $(1-\phi)$ in the relative variance is crucial for obtaining good collapse at high packing fractions.

\subsection{Vacancy Patterns}

Another type of disordered pattern can be created by randomly removing a fraction $f$ of particles from a crystalline lattice.  Here we study such vacancy patterns made from pixel particles on a two-dimensional triangular lattice that is rotated by $14^\circ$ with respect to the image grid.  The rotation, and the triangular pattern, help smooth out irregular features in the area fraction distributions that would be especially strong for a square lattice with square measuring windows.  We take the lattice spacing to be $b=30p_o$, which gives the area fraction of the base crystal as $\phi_{c}=\sqrt{4/3}(p_o/b)^2=0.00128$.  When a fraction $f$ of sites are vacant, the area fraction decreases to $\phi=(1-f)\phi_c$.  These are small compared to 1, so we may neglect factors of $(1-\phi)$ and our simulations are effectively in the continuum limit with Poisson statistics.

\begin{figure}[ht]
\includegraphics[width=2.75in]{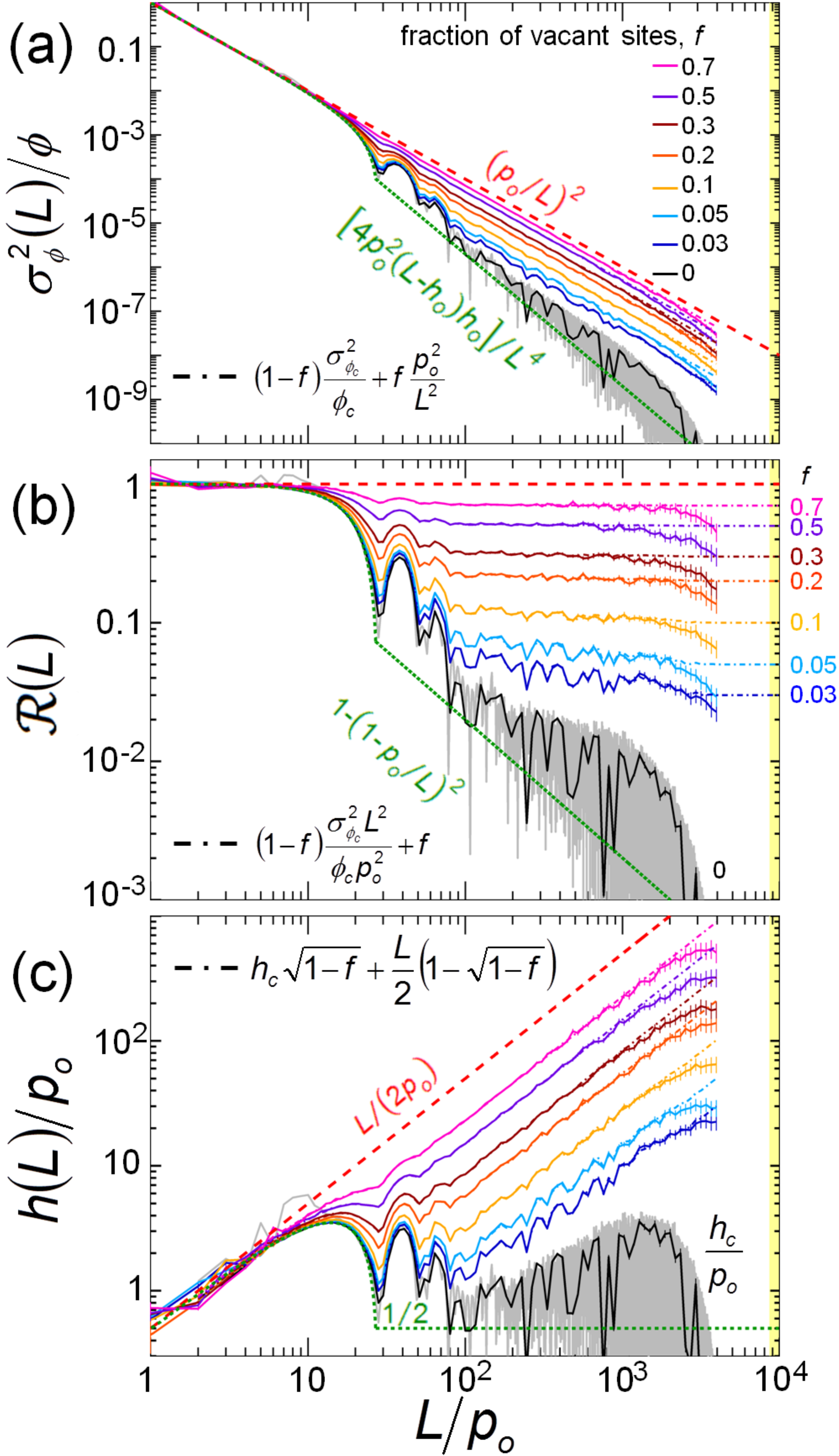}
\caption{(Color online) Area fraction variance (a), variance ratio (b), and hyperuniformity disorder length (c) versus measuring window size, for simulated two-dimensional triangular crystals with a specified fraction $f$ of random vacancies.  Here $\phi$ is the area fraction, $p_o$ is the pixel width, $8600p_o$ is image width (yellow shading), and $b=30p_o$ is the lattice spacing.  Measurements are made at all integer $L/p_o$ for $f=0$ (gray curve), but only at a select subset for $f>0$.  Results for $f=0$ at the same subset of $L/p_o$ are shown by black curves.  The dash-dotted curves correspond to our model, Eq.~(\ref{vvac}); these curves appear to stop below a certain $L$, but in fact they are present but entirely covered by the data curve.  The red and green dashed curves represent the upper and lower bounds given by Eqs.~(\ref{vbounds}-\ref{hbounds}).  The separated-particle bound is shown only for the crystal.}
\label{vacancies}
\end{figure}

Simulation results for the relative variance are shown in Fig.~\ref{vacancies}a for several values of $f$.  Each curve represents an average of 20 independent runs.  All data initially decay as $(p_o/L)^2$, just like for random patterns.  As $L$ increases toward the lattice spacing, the variance data fall below the $(p_o/L)^2$ upper bound, more quickly for smaller $f$.  This behavior matches well the lower bound given by the small-$L$ separated-particle expectation of Eq.~(\ref{vbounds}).  For $f>0$, the final decay is not as steep and appears to scale as $\sim1/L^2$.  At small $L$ there are prominent oscillations set by the lattice spacing.  For large $L$ and $f=0$, the data appear to eventually approach and hug the maximally-hyperuniform bound.  These trends are easier to see in terms of the corresponding variance ratios, $\mathcal R(L)$, plotted underneath in Fig.~\ref{vacancies}b.  In particular, it's more apparent that the initial decay is in good agreement with the small-$L$ separated-particles expectation of Eq.~(\ref{rbounds}).  And for large-$L$, before finite-size effects become strong, the $\mathcal R(L)$ data appear to approach a constant that equals the fraction $f$ of vacancies.  These features may all be readily understood, next, taking the relative variance of the perfect crystal as a given and using the binomial distribution to treat vacancies.

To model the volume fraction variance of vacancy patterns in $d$-dimensions, first note that the number of particles in a particular $L^d$ measuring window is equal to the number $N=N_c-N_v$ of enclosed crystal lattice sites minus the number of vacancies.  Therefore, the average over a large set of $L^d$ measuring windows is $\overline N = \overline{N}_c-\overline{N}_v$, and the variance is ${\sigma_N}^2 = {\sigma_{N_c}}^2 - 2( \overline{N_c N_v} - \overline{N}_c\overline{N}_v) + {\sigma_{N_v}}^2$.  For the crystal lattice sites, we may write $\overline{N}_c = \sum{n p_n}$ and $\overline{N_c^2} = \sum{n^2 p_n}$ where $p_n$ is the probability of finding $n$ sites in a randomly placed $L^d$ window.  If there are $n$ sites in a window, then the probability for $k$ of them to be vacant is given by the binomial distribution as $q_k = \{n!/[k!(n-k)!]\}f^k (1-f)^{n-k}$.  By direct summation, this gives $\overline{N}_v = \sum_n\sum_{k=0}^n(kq_k)p_n = f \overline{N}_c$ as expected.  Similar computation gives $\overline{N_v^2} = \sum_n\sum_{k=0}^n(k^2q_k)p_n = f(1-f)\overline{N}_c + f^2\overline{N_c^2}$ and $\overline{N_cN_v} = \sum_n\sum_{k=0}^n(np_n)(kq_k) = f\overline{N_c^2}$.  Plugging these into the expression for ${\sigma_N}^2$, multiplying by $(p_o/L)^{2d}$ to convert to area fraction variance, and dividing left and right hand sides by $\phi=(1-f)\phi_c$, gives a final result that is surprisingly simple:
\begin{equation}
	\frac{ {\sigma_\phi}^2(L) }{\phi} = (1-f)\frac{ {\sigma_{\phi_c}}^2(L) }{\phi_c} + f\left( \frac{p_o}{L} \right)^d. \label{vvac}
\end{equation}
Thus the relative variance of a vacancy pattern is the weighted average of that for a perfect crystal plus that for a random binomial pattern.  The variance ratio and hyperuniformity length can then be expressed in terms of those quantities for the perfect crystal as
\begin{eqnarray}
	\mathcal R(L) &=& (1-f)\mathcal R_c(L) + f, \label{rvac} \\
	h(L) &=& h_c(L)(1-f)^{1/d} + \frac{L}{2}[1-(1-f)^{1/d}]. \label{hvac}
\end{eqnarray}
Again these are weighted averages, but for $h(L)$ the weighting is not linear in $f$.  For $f\rightarrow 0$ the crystal results are recovered; and for $f\rightarrow 1$, as all particles are removed, the pattern becomes totally random with $h=L/2$.

Since crystals are hyperuniform, ${\sigma_\phi}^2(L)$ should decay faster than $1/L^d$ and therefore the second terms in Eqs.~(\ref{vvac}-\ref{hvac}) will eventually dominate at large $L$.  Then the large-$L$ asymptotic behavior is 
\begin{eqnarray}
	\frac{ {\sigma_\phi}^2(L)}{\phi} &=& f\left( \frac{p_o}{L} \right)^d, \label {vvacLargeL} \\
	\mathcal R(L) &=& f, \label{rvacLargeL} \\
	h(L) &=& (L/2)[1-(1-f)^{1/d}]. \label{hvacLargeL}
\end{eqnarray}
For our $d=2$ simulations, these predicted scalings with $L$ are readily seen in Fig.~\ref{vacancies}.  As a stronger test, $\mathcal V_c(L)$, $\mathcal R_c(L)$, and $h_c(L)$ were extracted from the $f=0$ data, and the resulting expectations based on Eqs.~(\ref{vvac}-\ref{hvac}) are plotted as dot-dashed curves.  These match the $f>0$ simulation data very well, until finite-size effects become strong.  At our level of precision, such effects are noticeable even for $L$ down to one tenth the sample width.

Based on the $\mathcal R(L)\rightarrow f$ result for vacancy patterns, we may interpret $\mathcal R(L)$ for general systems not just as a kind of randomness index but also more specifically as the fraction of space available for density fluctuations at wavelength $L$.

\subsection{Einstein Patterns}

Our previous examples all exhibited $h\sim L$ at long length scales, so now we study a type of disordered pattern that has no long-range density fluctuations and is unquestionably hyperuniform.  In particular, we consider two-dimensional pixel particles that are individually displaced by a Gaussian-distributed amount from a crystalline lattice.  We dub these ``Einstein patterns" in honor of Einstein's simplified model for the heat capacity of solids, where each atom is harmonically bound to a fixed lattice site.  This may be compared with the ``shuffled lattice'' patterns of Ref.~\cite{GabrielliPRD2002}, where each object is placed randomly inside a cubic volume surrounding a lattice site.  For our simulations we again use a triangular lattice with spacing $b=30p_o$, which is large enough that factors of $(1-\phi)$ may be dropped.  Displacements in each dimension are randomly drawn from a Gaussian distribution specified by the dimensionless parameter $\delta = x_{rms}/b = y_{rms}/b$.  For large $\delta$, it is sometimes necessary to repeat a trial placement until an unoccupied pixel is found.

\begin{figure}[ht]
\includegraphics[width=2.250in]{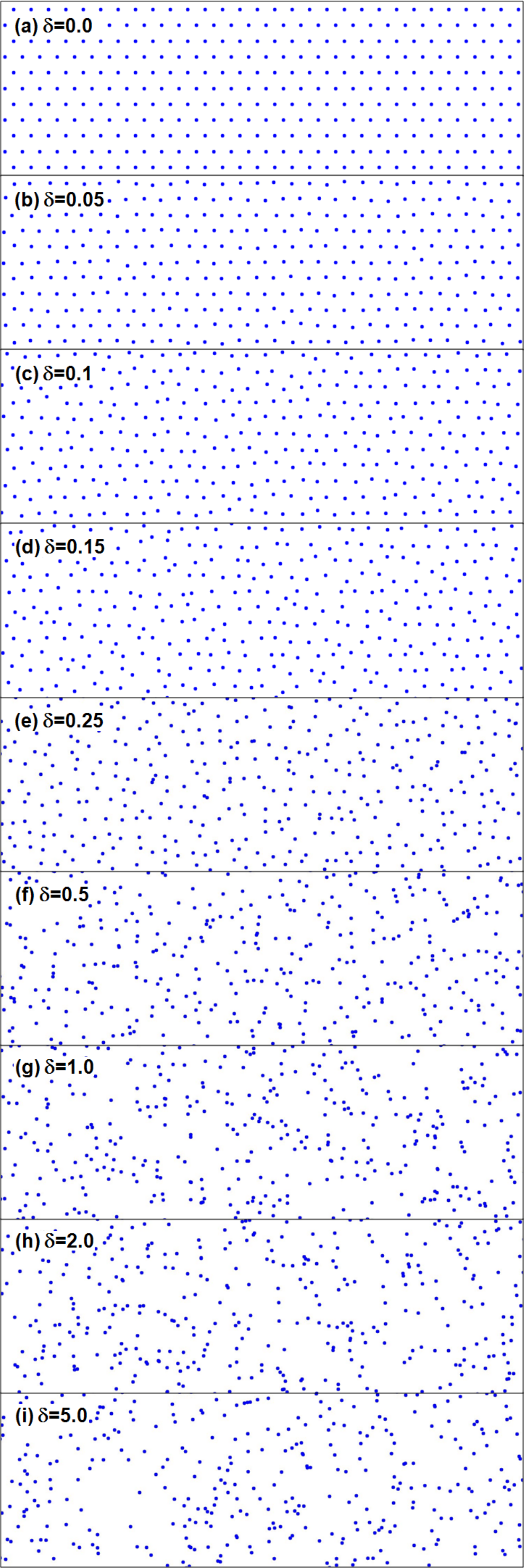}
\caption{(Color online) Example Einstein patterns for different root-mean-square displacements, labeled by the value of $\delta=x_{rms}/b=y_{rms}/b$.  Here the lattice spacing $b$ is 30 pixels, and for clarity the particles are shown as dots that are many pixels across.  By eye, patterns (f)-(i) are nearly indistinguishable and appear quite random.}
\label{meltpics}
\end{figure}

Example patterns are shown in Fig.~\ref{windowschematic} for one $\delta$, and in Fig.~\ref{meltpics} for a sequence of increasing $\delta$.  We include $\delta=0$, although the crystal isn't perfect due to pixilation effects that must be large for $x_{rms}<p_o$, i.e.\ here for $\delta<0.02$.  Note that the underlying crystal is evident for $\delta\ll1$, including the case $\delta=0.15$ that corresponds to the Lindemann criterions for melting.  For $\delta>0.5$, i.e.\ for displacements larger than about 1/2 lattice spacing, it is difficult to detect the lattice by eye.  Nevertheless, even for very large $\delta$, the particles are still bound to a lattice and hence cannot have long-range density fluctuations.  This will be reflected intuitively, as shown next, in terms of the behavior of the hyperuniformity disorder length.

\begin{figure}[ht]
\includegraphics[width=2.75in]{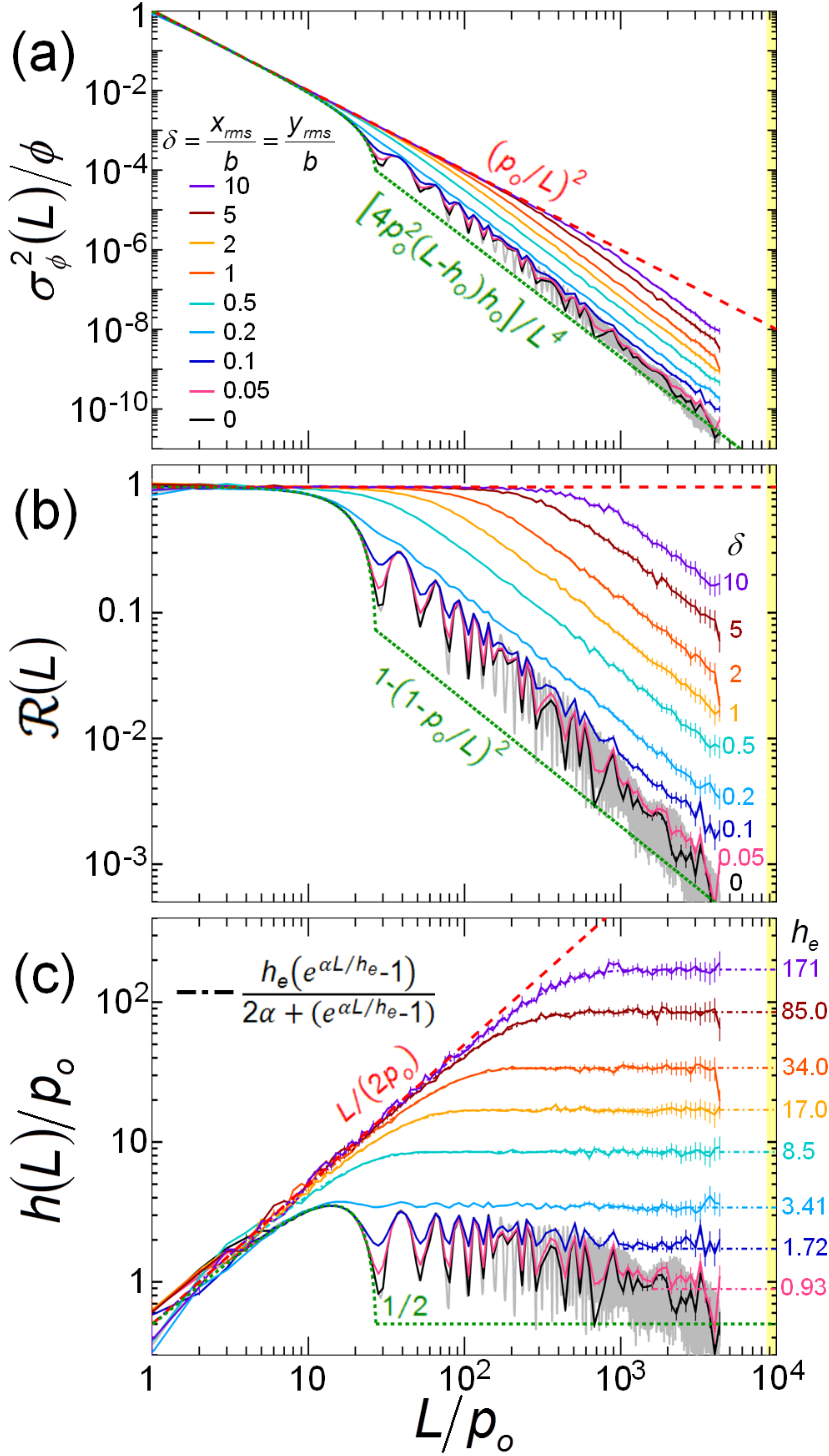}
\caption{(Color online) Area fraction variance (a), variance ratio (b), and hyperuniformity disorder length (c) versus measuring window size, for simulated two-dimensional Einstein patterns with specified dimensionless root-mean-square displacement, $\delta$.  Here $\phi$ is the area fraction, $p_o$ is the pixel width, $8600p_o$ is image width (yellow shading), and $b=30p_o$ is the lattice spacing.  In (c) an empirical fitting function is shown, which may be used to estimate the constant value $h_e$ of the hyperuniformity length at large $L$.  Note that $h_e$ is roughly one-half the root-mean-square displacement in each dimension.  Measurements are made at all integer $L/p_o$ for $\delta=0$ (gray curve), but only at a select subset for $\delta>0$.  Results for $\delta=0$ at the same subset of $L/p_o$ are shown by black curves.  The red and green dashed curves represent the upper and lower bounds given by Eqs.~(\ref{vbounds}-\ref{hbounds}).}
\label{einstein}
\end{figure}

We next measure the area fraction variance as above, and display the results in Fig.~\ref{einstein}a.  The corresponding variance ratios and hyperuniformity disorder lengths are shown underneath, in Figs.~\ref{einstein}b-c.  Each curve represents an average of fourteen Einstein patterns, all with different displacements, and where the underlying lattice was rotated by a different angle $\theta \in \{1^\circ, 2^\circ, \ldots, 14^\circ\}$.  This helps minimize artifacts from pixelation as well as from commensuration of the lattice with the square measuring windows.  Such artifacts are still present for the ``zero-temperature'' crystal case, $\delta=0$, where the relative variance is computed at each integer $L/p_o$ in the range $1-4300$.   Since these artifacts vanish for large enough $\delta$,  we then measure at fewer window sizes.

The observed behavior is as follows.  At small $L$, the relative variances for all $\delta$ match well with the random binomial expectation with $h=L/2$.   For increasing $L$, each data curve eventually falls below the binomial expectation -- sooner for smaller $\delta$, i.e\ for less disorder.  The $\delta=0$ crystal results show pronounced oscillations with features located according to lattice spacing.   At large enough $L$ such features are less regular, though are still present, and the variance approaches the lower bound of maximally hyperuniform.   For $\delta>0$, the variance results similarly show a final decay of $1/L^3$ but with a numerical prefactor that grows with $\delta$.   Einstein patterns are therefore all hyperuniform, but not maximally so.  

\begin{figure}[ht]
\includegraphics[width=2.75in]{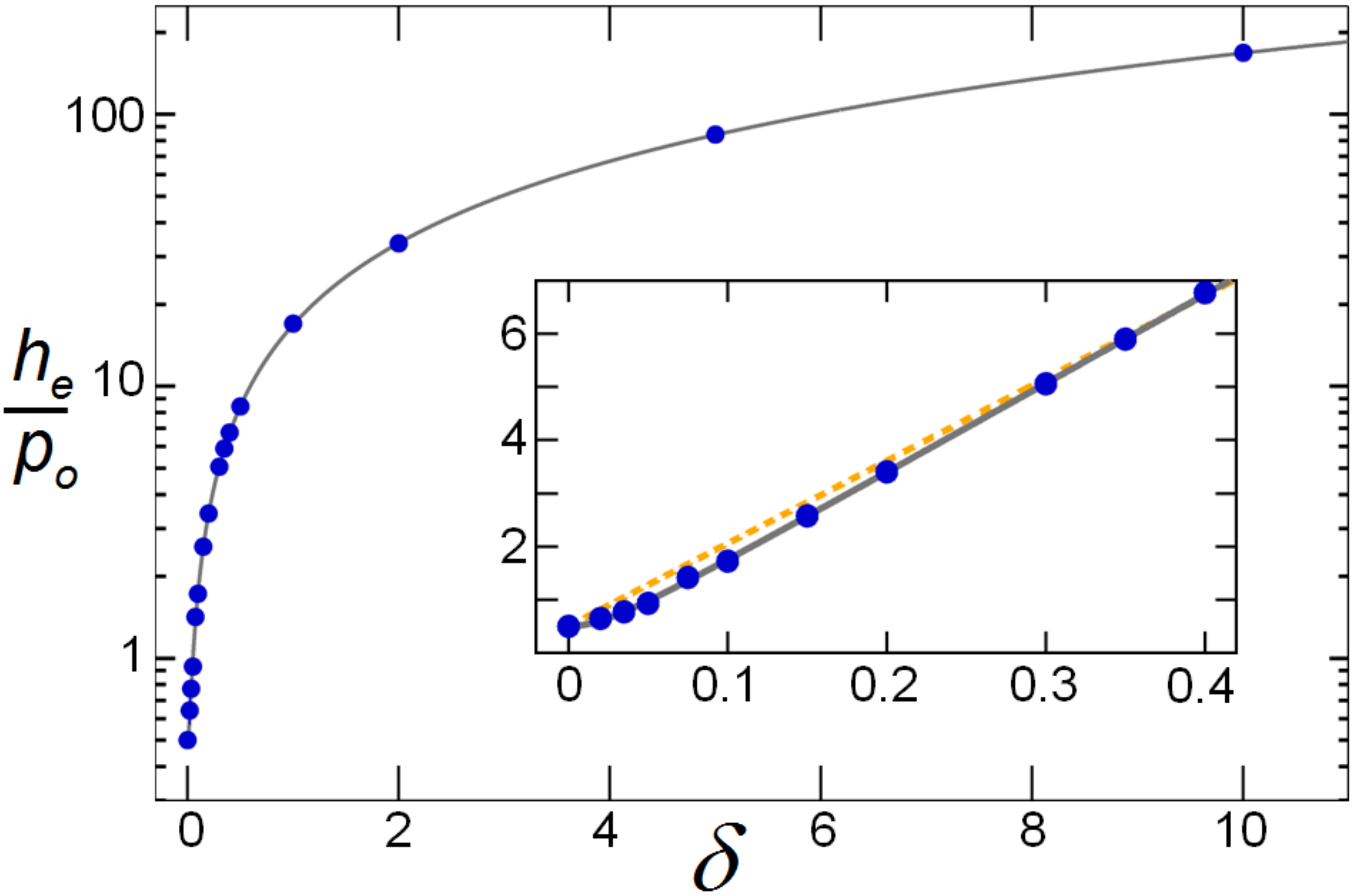}
\caption{(Color online) Large-$L$ value of the hyperuniformity disorder length versus dimensionless root-mean square displacement, $\delta=x_{rms}/b=y_{rms}/b$, for the Einstein pattern results shown in Fig.~\protect{\ref{einstein}}b with lattice spacing $b=30p_o$.  The insert is a blow-up of the small-$\delta$ region.  The fitting function $h_e = \sqrt{ {(p_o/2)}^2 + (c b \delta)^2 }$ matches the data well with $c=0.56\pm0.01$.  By contrast a linear fit for all $\delta$, shown by the dashed yellow line in the inset, fails for small $\delta$.}
\label{he}
\end{figure}

The extent of hyperuniformity may be diagnosed in terms of the behavior seen in Fig.~\ref{einstein}c for the hyperuniformity disorder length at large $L$.  In particular a signature of strong hyperuniformity is that $h$ becomes a constant, and the degree of hyperuniformity may be judged by the size $h_e$ of this constant.  To deduce $h_e$ we either average the large-$L$ results or we fit to
\begin{equation}
	h(L) = \frac{ h_e ( e^{\alpha L/h_e}-1) }{ 2\alpha+(e^{\alpha L/h_e}-1)   },
\label{hu_einstein}
\end{equation}
where $\alpha$ and $h_e$ are adjustable parameters.  This empirical function has an exponential crossover from $h=L/2$ at small $L$ to $h=h_e$ at large $L$.  It is seen in Fig.~\ref{einstein}c to match the data quite well, except for the oscillations coming from the crystal.  The fitting results for $\alpha$ are constant: $\alpha=0.77\pm0.02$.  The fitting results for $h_e$ are plotted versus $\delta=x_{rms}/b=y_{rms}/b$ in Fig.~\ref{he}, and are well described by
\begin{equation}
	h_e = \sqrt{ {(p_o/2)}^2 + (c x_{rms})^2 },
\label{hefit}
\end{equation}
with $c \approx 1/2$.  Thus, $h_e$ is approximately one-half the total root-mean-square displacement in each dimension, which comes from the combination of pixelation and Gaussian kicks.  This makes intuitive sense by thinking of a fixed measuring window and harmonically bound particles that independently oscillate in time with thermal energy.  Number fluctuations are primarily due to the particles whose lattice sites lie within a distance $h_e$ from the boundary of the measuring window.  This degree of disorder is not evident to the eye when the root-mean-square displacement is larger than the lattice spacing.  But it is readily detected by analyzing the relative variance in terms of the hyperuniformity disorder length.

\subsection{Finite-size effects}

\begin{figure}[ht]
\includegraphics[width=2.75in]{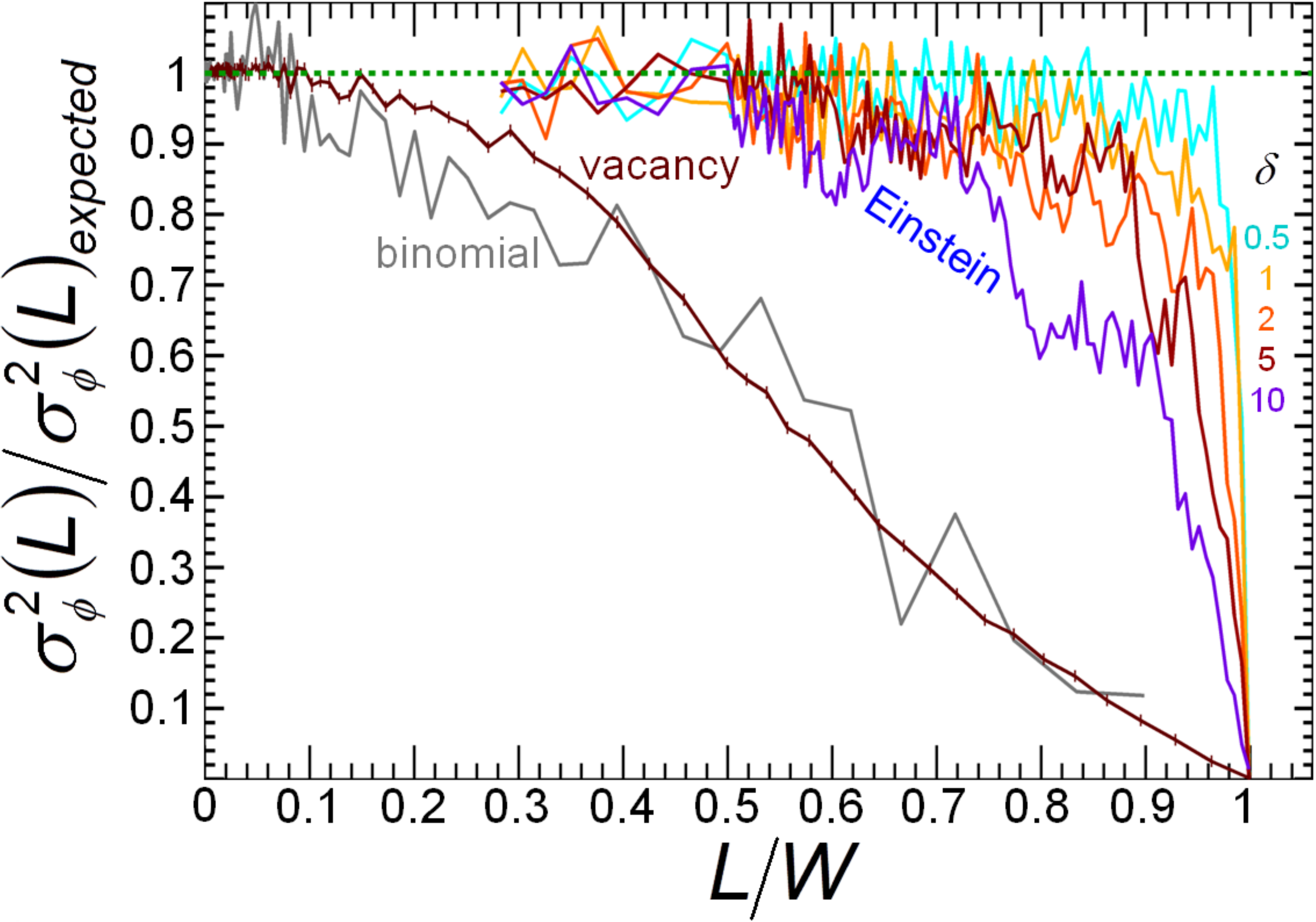}
\caption{(Color online) Ratio of simulated to expected area fraction variance versus $L/W$ where $L$ is the width of the measuring boxes and $W=8300p_o$ is the width of the sample.  Data are from previous figures.  As seen, the effect of finite sample size is to suppressed the variance at large $L$.   The longer the range of the density fluctuations, the larger the effect.
}
\label{finitesize}
\end{figure}

The degree to which variance measurements are affected by the finite size of the system can be examined in terms of the ratio of measured to expected variance as a function of $L/W$ where $W$ is the width of the sample.  This is shown in Fig.~\ref{finitesize} for all three pattern types discussed above.  For the binomial and vacancy patterns, we found no trend with area fraction and $f$, respectively.  So results are averaged together.  As seen in Fig.~\ref{finitesize}, finite size effects are generally largest for the binomial patterns and smallest for the Einstein patterns.  In all cases, the variance is suppressed and the amount increases with disorder and the range of the density fluctuations.  Thus Einstein patterns with small $\delta$ are least affected, and binomial patterns are most affected.   For general guidance, systematic errors are no more than about 1\% for $L<0.05W$ and no more than about 10\% for $L<0.2W$.  But for strongly hyperuniform systems the systematic errors can be considerably less.

\section{Grayscale Pixel Particles}

In this section we apply the same analysis approach to three different types of simulated two-dimensional patterns where the pixel particles are now polydisperse, with a range of intensity values $I_i$ that differ from one.  The particle areas are $a_i=I_i {p_o}^2$, and the expectations are determined by the weighted average $\langle a\rangle = \langle I \rangle {p_o}^2 = \sum \phi_i a_i / \phi$.   Area fractions are given by the sum of intensities divided by number of pixels, and the variance ${\sigma_\phi}^2(L)$ is measured using $L\times L$ square windows.

\subsection{Random Multinomial Patterns}

\begin{figure}[ht]
\includegraphics[width=2.75in]{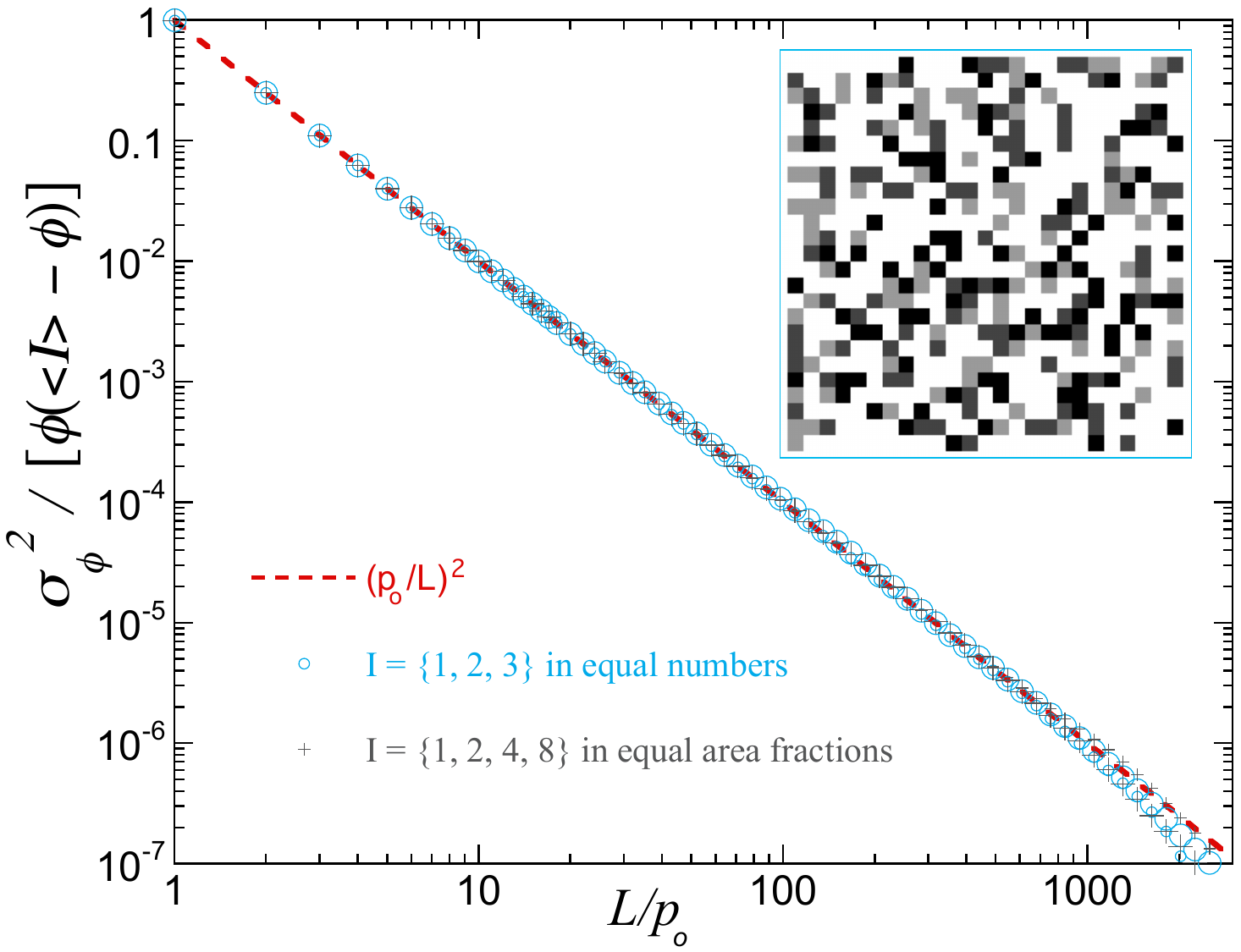}
\caption{(Color online) Normalized area fraction variance versus measuring window size for four random multinomial pixel patterns.  Two mixtures of pixel particle species are used, as labeled, and patterns for each are created with total area fractions of 0.1 (smaller symbols) and 0.9 (larger symbols).  A small portion of the $\{1,2,3\}$ system with $\phi=0.9$ is shown in the inset.  For all four patterns, the relative variance results agree well with the $(p_o/L)^2$ expectation shown by the dashed red line.}
\label{multi}
\end{figure}

The first test is for totally random multinomial patterns made either from equal {\it numbers} of three species of pixel particle with grayscale intensities $\{1, 2, 3\}$, or for patterns made from equal {\it area fractions} of four species of pixel particle with grayscale intensities $\{1, 2, 4, 8\}$.  Two patterns are created for each mixture, with different total area fractions of 0.1 and 0.9, by randomly choosing pixels values as $\{0, I_1, I_2,\ldots\}$ with appropriate probabilities $\{1-\sum q_i, q_1, q_2,\ldots\}$.  The relative variance is expected to be $\mathcal V(L) = {\sigma_\phi}^2(L)/[\phi(1-\phi/\langle I\rangle] = \langle I\rangle (p_o/L)^2$, where the $\phi/\langle I \rangle$ term cannot be neglected.  We therefore plot $\mathcal V(L)/\langle I \rangle$ in Fig.~\ref{multi} and compare with $(p_o/L)^2$.  As expected, the data for all four patterns perfectly collapse to this power law until finite size effects become noticeable at large $L$.  This validates the multinomial statistics calculation, and serves to emphasize that the relevant average particle area is set by $\phi_i$-weighting.


\subsection{Bidisperse Squares: Overlapping}

\begin{figure}[ht]
\includegraphics[width=2.75in]{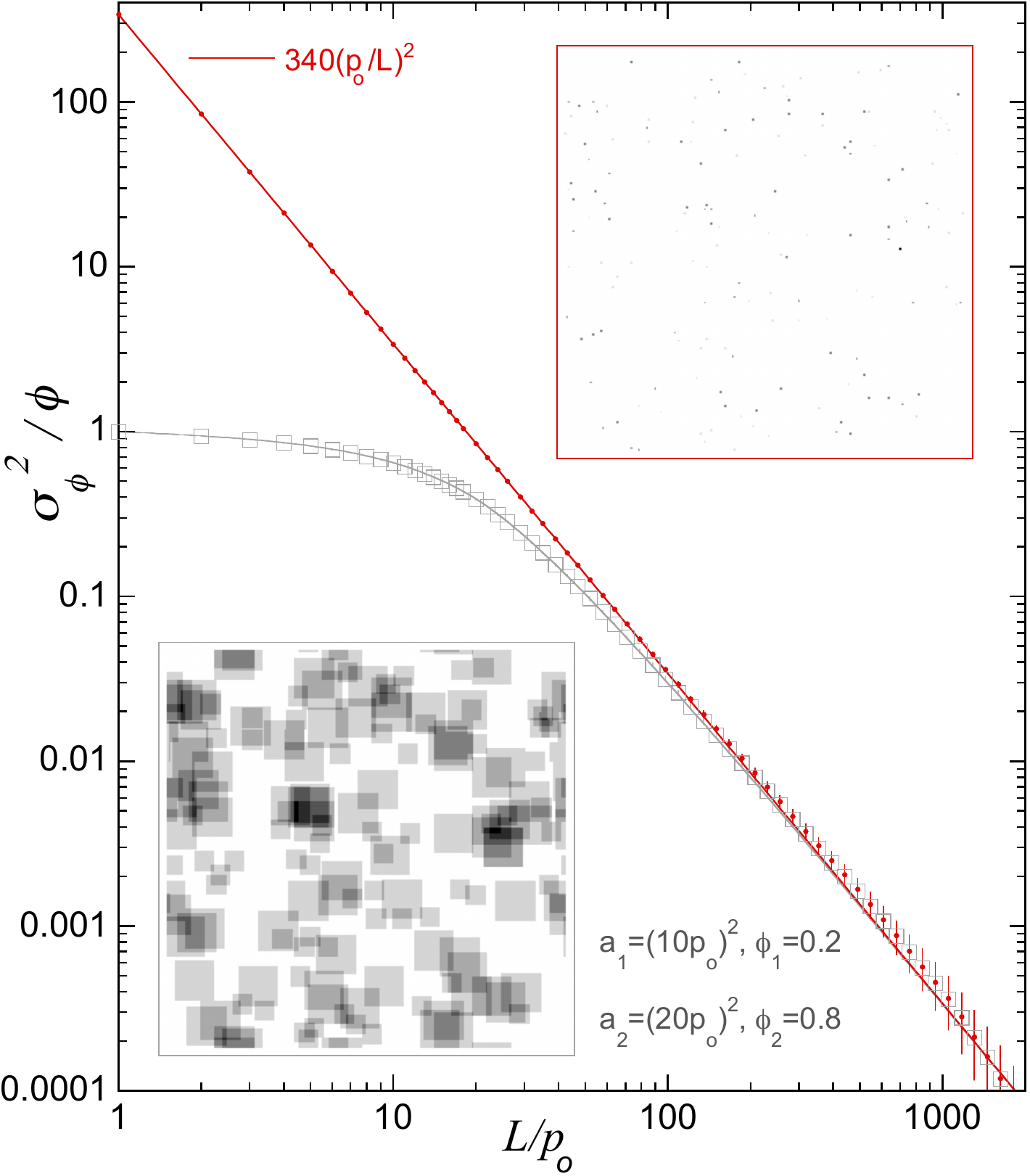}
\caption{(Color online) Relative variance versus measuring window size for a random arrangement of a 50:50 bidisperse mixture of extended square particles (gray squares) and the corresponding central-pixel representation (red dots).  The actual image sizes are $4300p_o\times4300p_o$; much smaller versions are shown as insets.  Particle sizes and total area fractions are labeled.  The $\phi_i$-weighted average area of the two species is $\langle a\rangle=340{p_o}^2$.  The red line is the $\langle a\rangle /L^2$ expectation.  The gray curve is from Ref.~\protect{\cite{DJDhudls}}.}
\label{bipoiss}
\end{figure}

The second test is for a grayscale Poisson pattern made from of a 50:50 mixture of two different size square particles, $a_1=(10p_o)^2$ and $a_2=(20p_o)^2$, and the associated central pixel pattern.  Extended particles are repeatedly added at random locations, such that each covered pixel is incremented by +1, until the respective area fractions reach $\phi_1=0.2$ and $\phi_2=0.8$.  The total area fraction is then $\phi=1$ and the $\phi_i$-weighted average particle area is $\langle a\rangle = 340{p_o}^2$.  Note that particle-particle overlaps are freely allowed, and that the value of each pixel is equal to the number of particles that cover it.  In parallel we construct the central pixel pattern representation of the same configurations, where the center pixel for each particle of each species is incremented by $I_1=100$ or $I_2=400$.  For both image types, the local area fraction is given by the sum of pixel values per unit area.  Small examples of extended-particle and corresponding central-pixel images are shown in the insets of Fig.~\ref{bipoiss}.  Even though the total area fraction is one, the particles are large enough that the central pixel pattern is quite dilute and has intensity values of only $\{0, I_1, I_2\}$.  Therefore, the distinction between Poisson and multinomial statistics for the central pixel representation can be neglected.

The main plot of Fig.~\ref{bipoiss} shows the relative area fraction variance for the two different representations of the same arrangement.  The results for the central-pixel image are well described by the prediction ${\sigma_\phi}^2/\phi =\langle a\rangle / L^2 = 340(p_o/L)^2$.  By contrast, the relative variance for the extended particle image merges onto $\langle a\rangle / L^2$ only for $L$ much greater than the particle sizes; in this limit the two representations give identical results.  At small-$L$ the relative variance for the extended particle image has a slower decay starting from ${\sigma_\phi}^2/\phi=1$ at $L=p_o$.  This reflects the particle shape as well as the random arrangement \cite{DJDhudls}, whereas the decay for central pixel patterns reflects only the nature of the arrangement.

\subsection{Bidisperse Squares: Non-Overlapping}

\begin{figure}[ht]
\includegraphics[width=2.75in]{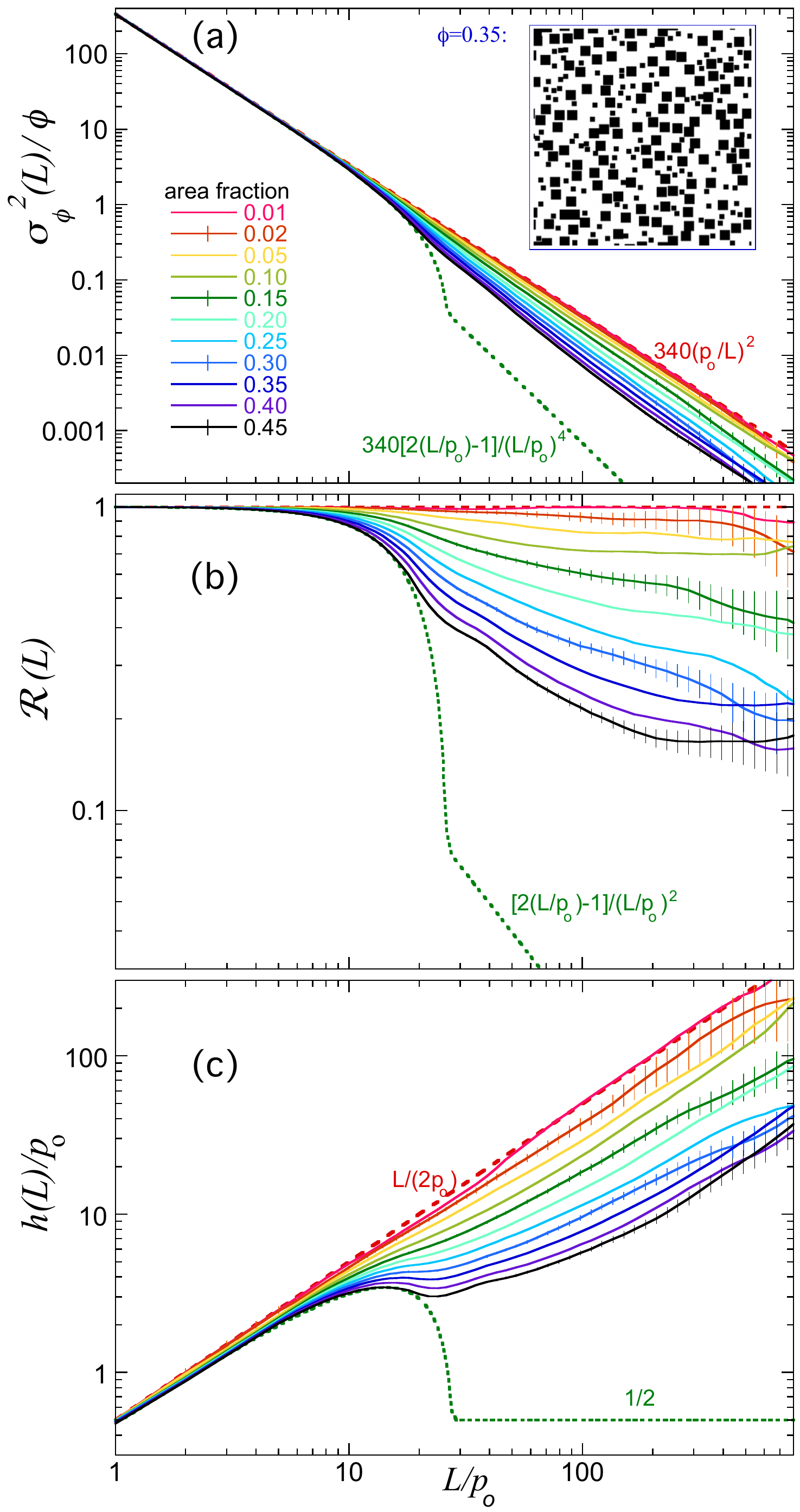}
\caption{(Color online) Relative variance (a), variance ratio (b), and hyperuniformity disorder length (c) versus measuring window size for the central-pixel representation of arrangements of non-overlapping $a_1=(10p_o)^2$ and $a_2=(20p_o)^2$ square particles, in equal numbers, like in the inset, for several area fractions.  For clarity, error bars are plotted on only a few data as sets marked in the legend.  The red and green broken curves represent the upper and lower bounds given by Eqs.~(\ref{vbounds}-\ref{hbounds}).  The separated-particle bound is shown only for the largest packing fraction.}
\label{bibino}
\end{figure}

Lastly, we use a central-pixel representation to diagnose the uniformity of simulated 50:50 bidisperse mixtures of $a_1=(10p_o)^2$ and $a_2=(20p_o)^2$ square particles with $\langle a\rangle = 340{p_o}^2$ as a function of total area fraction.  This system is like in Fig.~\ref{bipoiss}, but with one major difference:  Random trial locations are now rejected if any particle-particle overlap occurs.  An example of the resulting binary pattern of extended particles is shown in the inset of Fig.~\ref{bibino}.  As before, central-pixel representations are simultaneously made and then analyzed.

Results for the relative variance, the variance ratio, and the corresponding hyperuniformity disorder lengths, are collected in Fig.~\ref{bibino}.  The general behavior is an amalgam of key features seen for the vacancy and Einstein patterns.  Like the latter, the system matches the separated-particles expectations at small $L$ followed by a developing plateau of constant $h(L)$ -- but only out to intermediate $L$.  At larger $L$, the scaling then crosses over to that seen for the vacancy patterns with long-range density fluctuations: $\mathcal V(L)\sim 1/L^2$, $\mathcal R(L)\sim$~constant, and $h(L)\sim L$.  With increasing $\phi$, the plateau becomes more pronounces and the long-range density fluctuations decrease.  If the length of the plateau were to diverge as $\phi$ increases toward random-close packing, then the onset of jamming would be accompanied by the development of strong hyperuniformity like in an Einstein pattern.  This is the topic of a following paper \cite{ATCjam}, where different packing protocols are used to generate bigger systems with higher packing fractions than are accessible here.  For now, Fig.~\ref{bibino} serves as demonstration of the HUDLS analysis method for polydisperse hard or soft particles and, hopefully, also whets the appetite for its larger scale uses.

\vfill\break

\section{Conclusion}

In summary, we have introduced a powerful way to analyze ${\sigma_\phi}^2(L)$ versus $L$ results and thereby diagnose the extent of hyperuniformity in terms of two real-space spectra:  (a) the ratio $\mathcal R(L)$ of the variance for the pattern in question to that for a random arrangement of the same set of particles, and (b) the  hyperuniformity disorder length $h(L)$.  Whereas $\mathcal R(L)$ is a kind of randomness index that can also be interpreted as the fraction of space available for density fluctuations at wavelength $L$, the value of $h(L)$ specifies the distance from the boundary of $V_\Omega = L^d$ measuring windows over which fluctuations are important.  This connects directly to the original idea that hyperuniform systems are controlled by particles on the surface of the measuring windows \cite{TorquatoPRE2003}, and it makes the large-$L$ scaling ${\sigma_\phi}^2(L)\propto \langle v\rangle h/L^{d+1}$ dimensionally correct.  This applies equally well to liquid-like systems having long-range density fluctuations with $\mathcal R(L)\rightarrow$~constant and $h(L)\sim L$, as well as to strongly hyperuniform systems with $\mathcal R(L)\sim 1/L$ and $h(L)\rightarrow$~constant, and to systems anywhere between.

Whereas prior work focuses on the form of the large-$L$ scaling for diagnosing hyperuniformity, an important feature of our work is to bring meaning to the {\it value} of the variance, though intuitive interpretation of the corresponding values of $\mathcal R(L)$ and $h(L)$.  These quantities are even further useful when compared to the exact bounds we computed for pixelated space, using Poisson and multinomial statistics.  The resulting formulae and plots are expressed using the pixel length, $p_o$, explicitly, so as to be dimensionally correct.  

While pixel patterns are natural for digital images coming from experiment, they might seem inappropriate for arrangements of {\it point} particles in continuous space.  In fact they literally apply via the realization that limitations on numerical precision effectively set $p_o$.  But furthermore, there is a well-defined continuum limit where the particle volume $v_i=I_i{p_o}^d$ remains constant while pixel length vanishes and the intensity diverges.  The relevant formulae for a point representation of extended particles with $\phi_i$-weighted average volume $\langle v\rangle$, and $d$-dimensional spherical measuring windows of volume $V_\Omega$ and diameter $D$, then become
\begin{eqnarray}
	\mathcal V(D) &\equiv& {\sigma_\phi}^2(D)/\phi \\
	\Delta \mathcal V &=&  \sqrt{  \frac{ \langle v^3\rangle /{ {V_\Omega}^3} }{ S\phi }  + \frac{2\mathcal V^2}{S-1}  } \\
	\mathcal R(D) &=& \frac{ \mathcal V(D)D^d }{  [ \mathcal V(D_o)+\phi]{D_o}^d  } \\
	1 - \phi V_\Omega/ \langle v \rangle &\le&  \mathcal R(D) \le 1 \\
	h(D) &\equiv& (D/2)\{ 1 - \left[  1 - \mathcal R(D) \right]^{1/d} \}
\end{eqnarray}
where $D_o$ is some smallest chosen measuring diameter that is less than the smallest particle-particle separation in the pattern, and $S$ is the number of independent samples made by the randomly chosen set of measuring windows.  The first and last of these are definitions;  the others are new results.  The upper and and lower bounds are respectively set by the calculated variance for the separated-particles and the totally-random arrangements.  The large-$L$ scalings are $\mathcal V(L)\sim 1/L^{d+\epsilon}$, $\mathcal R(L)\sim1/L^\epsilon$, and $h(L)\sim L^{1-\epsilon}$, where the exponent $\epsilon$ ranges from zero (liquid-like) to one (strongly hyperuniform).

Thus our general advances for analysis of particle arrangements include the concepts of $\mathcal R$ and $h$, their intuitive meanings and bounds, and a systematic way to estimate the expected statistical uncertainty -- for both pixelated and continuum space.  It is unclear how this translates or could be done in terms of the $S(q)$ or $\chi(q)$ spectra.  Therefore, we recommend that hyperuniformity be diagnosed from measurement of the real-space spectrum of ${\sigma_\phi}^2$ versus window size.  As standard procedure, essentially-raw variance data are to be plotted in terms of $\mathcal R\pm \Delta \mathcal R$ and then further analyzed in terms of $h\pm \Delta h$ versus measuring window size.

To verify expectations for random pixel patterns, and to build intuition and examine behavior relative to the bounds, we applied our Hyperuniformity Disorder Length Spectroscopy (HUDLS) approach to a variety of simulated two-dimensional patterns.  The results are exactly as expected for both the random binomial and multinomial patterns, as long as $L$ is small enough for finite-size effects to be small.  The results for the vacancy and the Einstein patterns can be simply understood, and capture essential features that emerge in a small-scale simulation of a bidisperse mixture of extended particles as the packing fraction is increased.  Thus the extent of hyperuniformity in polydisperse systems may now be diagnosed by similarly using HUDLS on larger-scale simulations and experiments.  This opens up new lines of research, including study of foams \cite{ATCfoam} as well as the jamming transition \cite{ATCjam}, for both of which polydispersity is often present and acts to suppress crystallization.

\begin{acknowledgments}
We thank Jim Sethna for suggesting a crystal with vacancies as a model system exhibiting long-range Poissonian density fluctuations.  This work was supported by NASA grant NNX14AM99G (primary) and by NSF grants MRSEC/DMR-1120901 and DMR-1305199.
\end{acknowledgments}

\bibliography{../HyperRefs}

\begin{thebibliography}{31}
\expandafter\ifx\csname natexlab\endcsname\relax\def\natexlab#1{#1}\fi
\expandafter\ifx\csname bibnamefont\endcsname\relax
  \def\bibnamefont#1{#1}\fi
\expandafter\ifx\csname bibfnamefont\endcsname\relax
  \def\bibfnamefont#1{#1}\fi
\expandafter\ifx\csname citenamefont\endcsname\relax
  \def\citenamefont#1{#1}\fi
\expandafter\ifx\csname url\endcsname\relax
  \def\url#1{\texttt{#1}}\fi
\expandafter\ifx\csname urlprefix\endcsname\relax\def\urlprefix{URL }\fi
\providecommand{\bibinfo}[2]{#2}
\providecommand{\eprint}[2][]{\url{#2}}

\bibitem[{\citenamefont{Torquato and Stillinger}(2003)}]{TorquatoPRE2003}
\bibinfo{author}{\bibfnamefont{S.}~\bibnamefont{Torquato}} \bibnamefont{and}
  \bibinfo{author}{\bibfnamefont{F.~H.} \bibnamefont{Stillinger}},
  \bibinfo{journal}{Phys. Rev. E} \textbf{\bibinfo{volume}{68}},
  \bibinfo{pages}{041113} (\bibinfo{year}{2003}).

\bibitem[{\citenamefont{Zachary and Torquato}(2009)}]{ZacharyJSM2009}
\bibinfo{author}{\bibfnamefont{C.~E.} \bibnamefont{Zachary}} \bibnamefont{and}
  \bibinfo{author}{\bibfnamefont{S.}~\bibnamefont{Torquato}},
  \bibinfo{journal}{J. Stat. Mech.: Theory and Experiment} p.
  \bibinfo{pages}{P12015} (\bibinfo{year}{2009}).

\bibitem[{\citenamefont{Donev et~al.}(2005)\citenamefont{Donev, Stillinger, and
  Torquato}}]{DonevPRL2005}
\bibinfo{author}{\bibfnamefont{A.}~\bibnamefont{Donev}},
  \bibinfo{author}{\bibfnamefont{F.~H.} \bibnamefont{Stillinger}},
  \bibnamefont{and} \bibinfo{author}{\bibfnamefont{S.}~\bibnamefont{Torquato}},
  \bibinfo{journal}{Phys. Rev. Lett.} \textbf{\bibinfo{volume}{95}},
  \bibinfo{pages}{090604} (\bibinfo{year}{2005}).

\bibitem[{\citenamefont{Torquato and
  Stillinger}(2010)}]{TorquatoStillingerRMP10}
\bibinfo{author}{\bibfnamefont{S.}~\bibnamefont{Torquato}} \bibnamefont{and}
  \bibinfo{author}{\bibfnamefont{F.~H.} \bibnamefont{Stillinger}},
  \bibinfo{journal}{Rev. Mod. Phys.} \textbf{\bibinfo{volume}{82}},
  \bibinfo{pages}{2633} (\bibinfo{year}{2010}).

\bibitem[{\citenamefont{Kurita and Weeks}(2010)}]{KuritaPRE2010}
\bibinfo{author}{\bibfnamefont{R.}~\bibnamefont{Kurita}} \bibnamefont{and}
  \bibinfo{author}{\bibfnamefont{E.~R.} \bibnamefont{Weeks}},
  \bibinfo{journal}{Phys. Rev. E} \textbf{\bibinfo{volume}{82}},
  \bibinfo{pages}{011403} (\bibinfo{year}{2010}).

\bibitem[{\citenamefont{Xu and Ching}(2010)}]{XuSM2010}
\bibinfo{author}{\bibfnamefont{N.}~\bibnamefont{Xu}} \bibnamefont{and}
  \bibinfo{author}{\bibfnamefont{E.~S.~C.} \bibnamefont{Ching}},
  \bibinfo{journal}{Soft Matter} \textbf{\bibinfo{volume}{6}},
  \bibinfo{pages}{2944} (\bibinfo{year}{2010}).

\bibitem[{\citenamefont{Berthier et~al.}(2011)\citenamefont{Berthier,
  Chaudhuri, Coulais, Dauchot, and Sollich}}]{BerthierPRL2011}
\bibinfo{author}{\bibfnamefont{L.}~\bibnamefont{Berthier}},
  \bibinfo{author}{\bibfnamefont{P.}~\bibnamefont{Chaudhuri}},
  \bibinfo{author}{\bibfnamefont{C.}~\bibnamefont{Coulais}},
  \bibinfo{author}{\bibfnamefont{O.}~\bibnamefont{Dauchot}}, \bibnamefont{and}
  \bibinfo{author}{\bibfnamefont{P.}~\bibnamefont{Sollich}},
  \bibinfo{journal}{Phys. Rev. Lett.} \textbf{\bibinfo{volume}{106}},
  \bibinfo{pages}{120601} (\bibinfo{year}{2011}).

\bibitem[{\citenamefont{Zachary et~al.}(2011)\citenamefont{Zachary, Jiao, and
  Torquato}}]{ZacharyPRL2011}
\bibinfo{author}{\bibfnamefont{C.~E.} \bibnamefont{Zachary}},
  \bibinfo{author}{\bibfnamefont{Y.}~\bibnamefont{Jiao}}, \bibnamefont{and}
  \bibinfo{author}{\bibfnamefont{S.}~\bibnamefont{Torquato}},
  \bibinfo{journal}{Phys. Rev. Lett.} \textbf{\bibinfo{volume}{106}},
  \bibinfo{pages}{178001} (\bibinfo{year}{2011}).

\bibitem[{\citenamefont{Kurita and Weeks}(2011)}]{KuritaPRE2011}
\bibinfo{author}{\bibfnamefont{R.}~\bibnamefont{Kurita}} \bibnamefont{and}
  \bibinfo{author}{\bibfnamefont{E.~R.} \bibnamefont{Weeks}},
  \bibinfo{journal}{Phys. Rev. E} \textbf{\bibinfo{volume}{84}},
  \bibinfo{pages}{030401} (\bibinfo{year}{2011}).

\bibitem[{\citenamefont{Ikeda and Berthier}(2015)}]{IkedaPRE2015}
\bibinfo{author}{\bibfnamefont{A.}~\bibnamefont{Ikeda}} \bibnamefont{and}
  \bibinfo{author}{\bibfnamefont{L.}~\bibnamefont{Berthier}},
  \bibinfo{journal}{Phys. Rev. E} \textbf{\bibinfo{volume}{92}},
  \bibinfo{pages}{012309} (\bibinfo{year}{2015}).

\bibitem[{\citenamefont{Dreyfus et~al.}(2015)\citenamefont{Dreyfus, Xu, Still,
  Hough, Yodh, and Torquato}}]{DreyfusPRE2015}
\bibinfo{author}{\bibfnamefont{R.}~\bibnamefont{Dreyfus}},
  \bibinfo{author}{\bibfnamefont{Y.}~\bibnamefont{Xu}},
  \bibinfo{author}{\bibfnamefont{T.}~\bibnamefont{Still}},
  \bibinfo{author}{\bibfnamefont{L.~A.} \bibnamefont{Hough}},
  \bibinfo{author}{\bibfnamefont{A.~G.} \bibnamefont{Yodh}}, \bibnamefont{and}
  \bibinfo{author}{\bibfnamefont{S.}~\bibnamefont{Torquato}},
  \bibinfo{journal}{Phys. Rev. E} \textbf{\bibinfo{volume}{91}},
  \bibinfo{pages}{012302} (\bibinfo{year}{2015}).

\bibitem[{\citenamefont{Wu et~al.}(2015)\citenamefont{Wu, Olsson, and
  Teitel}}]{WuPRE2015}
\bibinfo{author}{\bibfnamefont{Y.}~\bibnamefont{Wu}},
  \bibinfo{author}{\bibfnamefont{P.}~\bibnamefont{Olsson}}, \bibnamefont{and}
  \bibinfo{author}{\bibfnamefont{S.}~\bibnamefont{Teitel}},
  \bibinfo{journal}{Phys. Rev. E} \textbf{\bibinfo{volume}{92}},
  \bibinfo{pages}{052206} (\bibinfo{year}{2015}).

\bibitem[{\citenamefont{Florescu et~al.}(2009)\citenamefont{Florescu, Torquato,
  and Steinhardt}}]{FlorescuPNAS2009}
\bibinfo{author}{\bibfnamefont{M.}~\bibnamefont{Florescu}},
  \bibinfo{author}{\bibfnamefont{S.}~\bibnamefont{Torquato}}, \bibnamefont{and}
  \bibinfo{author}{\bibfnamefont{P.~J.} \bibnamefont{Steinhardt}},
  \bibinfo{journal}{Proc. Nat. Acad. Sci.} \textbf{\bibinfo{volume}{106}},
  \bibinfo{pages}{20658} (\bibinfo{year}{2009}).

\bibitem[{\citenamefont{Man et~al.}(2013)\citenamefont{Man, Florescu,
  Williamson, He, Hashemizad, Leung, Liner, Torquato, Chaikin, and
  Steinhardt}}]{ManPNAS2013}
\bibinfo{author}{\bibfnamefont{W.}~\bibnamefont{Man}},
  \bibinfo{author}{\bibfnamefont{M.}~\bibnamefont{Florescu}},
  \bibinfo{author}{\bibfnamefont{E.~P.} \bibnamefont{Williamson}},
  \bibinfo{author}{\bibfnamefont{Y.}~\bibnamefont{He}},
  \bibinfo{author}{\bibfnamefont{S.~R.} \bibnamefont{Hashemizad}},
  \bibinfo{author}{\bibfnamefont{B.~Y.~C.} \bibnamefont{Leung}},
  \bibinfo{author}{\bibfnamefont{D.~R.} \bibnamefont{Liner}},
  \bibinfo{author}{\bibfnamefont{S.}~\bibnamefont{Torquato}},
  \bibinfo{author}{\bibfnamefont{P.~M.} \bibnamefont{Chaikin}},
  \bibnamefont{and} \bibinfo{author}{\bibfnamefont{P.~J.}
  \bibnamefont{Steinhardt}}, \bibinfo{journal}{Proc. Nat. Acad. Sci.}
  \textbf{\bibinfo{volume}{110}}, \bibinfo{pages}{15886}
  (\bibinfo{year}{2013}).

\bibitem[{\citenamefont{Muller et~al.}(2014)\citenamefont{Muller, Haberko,
  Marichy, and Scheffold}}]{MullerAOM2014}
\bibinfo{author}{\bibfnamefont{N.}~\bibnamefont{Muller}},
  \bibinfo{author}{\bibfnamefont{J.}~\bibnamefont{Haberko}},
  \bibinfo{author}{\bibfnamefont{C.}~\bibnamefont{Marichy}}, \bibnamefont{and}
  \bibinfo{author}{\bibfnamefont{F.}~\bibnamefont{Scheffold}},
  \bibinfo{journal}{Adv. Optical Mater.} \textbf{\bibinfo{volume}{2}},
  \bibinfo{pages}{115} (\bibinfo{year}{2014}).

\bibitem[{\citenamefont{Ni et~al.}(2016)\citenamefont{Ni, Zhang, Qi, Yang,
  Chen, and Man}}]{ManOE16}
\bibinfo{author}{\bibfnamefont{P.}~\bibnamefont{Ni}},
  \bibinfo{author}{\bibfnamefont{P.}~\bibnamefont{Zhang}},
  \bibinfo{author}{\bibfnamefont{X.}~\bibnamefont{Qi}},
  \bibinfo{author}{\bibfnamefont{J.}~\bibnamefont{Yang}},
  \bibinfo{author}{\bibfnamefont{Z.}~\bibnamefont{Chen}}, \bibnamefont{and}
  \bibinfo{author}{\bibfnamefont{W.}~\bibnamefont{Man}},
  \bibinfo{journal}{Optics Express} \textbf{\bibinfo{volume}{24}},
  \bibinfo{pages}{2420} (\bibinfo{year}{2016}).

\bibitem[{\citenamefont{Froufe-P\'erez
  et~al.}(2016)\citenamefont{Froufe-P\'erez, Engel, Damasceno, Muller, Haberko,
  Glotzer, and Scheffold}}]{Scheffold2016}
\bibinfo{author}{\bibfnamefont{L.~S.} \bibnamefont{Froufe-P\'erez}},
  \bibinfo{author}{\bibfnamefont{M.}~\bibnamefont{Engel}},
  \bibinfo{author}{\bibfnamefont{P.~F.} \bibnamefont{Damasceno}},
  \bibinfo{author}{\bibfnamefont{N.}~\bibnamefont{Muller}},
  \bibinfo{author}{\bibfnamefont{J.}~\bibnamefont{Haberko}},
  \bibinfo{author}{\bibfnamefont{S.~C.} \bibnamefont{Glotzer}},
  \bibnamefont{and}
  \bibinfo{author}{\bibfnamefont{F.}~\bibnamefont{Scheffold}},
  \bibinfo{journal}{Phys. Rev. Lett.} \textbf{\bibinfo{volume}{117}},
  \bibinfo{pages}{053902} (\bibinfo{year}{2016}).

\bibitem[{\citenamefont{Jiao et~al.}(2014)\citenamefont{Jiao, Lau, Hatzikirou,
  Meyer-Hermann, Corbo, and Torquato}}]{JiaoPRE2014}
\bibinfo{author}{\bibfnamefont{Y.}~\bibnamefont{Jiao}},
  \bibinfo{author}{\bibfnamefont{T.}~\bibnamefont{Lau}},
  \bibinfo{author}{\bibfnamefont{H.}~\bibnamefont{Hatzikirou}},
  \bibinfo{author}{\bibfnamefont{M.}~\bibnamefont{Meyer-Hermann}},
  \bibinfo{author}{\bibfnamefont{J.~C.} \bibnamefont{Corbo}}, \bibnamefont{and}
  \bibinfo{author}{\bibfnamefont{S.}~\bibnamefont{Torquato}},
  \bibinfo{journal}{Phys. Rev. E} \textbf{\bibinfo{volume}{89}},
  \bibinfo{pages}{022721} (\bibinfo{year}{2014}).

\bibitem[{\citenamefont{Gabrielli et~al.}(2002)\citenamefont{Gabrielli, Joyce,
  and Labini}}]{GabrielliPRD2002}
\bibinfo{author}{\bibfnamefont{A.}~\bibnamefont{Gabrielli}},
  \bibinfo{author}{\bibfnamefont{M.}~\bibnamefont{Joyce}}, \bibnamefont{and}
  \bibinfo{author}{\bibfnamefont{F.~S.} \bibnamefont{Labini}},
  \bibinfo{journal}{Phys. Rev. D} \textbf{\bibinfo{volume}{65}},
  \bibinfo{pages}{083523} (\bibinfo{year}{2002}).

\bibitem[{\citenamefont{Hexner and Levine}(2015)}]{HexnerPRL2015}
\bibinfo{author}{\bibfnamefont{D.}~\bibnamefont{Hexner}} \bibnamefont{and}
  \bibinfo{author}{\bibfnamefont{D.}~\bibnamefont{Levine}},
  \bibinfo{journal}{Phys. Rev. Lett.} \textbf{\bibinfo{volume}{114}},
  \bibinfo{pages}{110602} (\bibinfo{year}{2015}).

\bibitem[{\citenamefont{Tjhung and Berthier}(2015)}]{TjhunPRL2015}
\bibinfo{author}{\bibfnamefont{E.}~\bibnamefont{Tjhung}} \bibnamefont{and}
  \bibinfo{author}{\bibfnamefont{L.}~\bibnamefont{Berthier}},
  \bibinfo{journal}{Phys. Rev. Lett.} \textbf{\bibinfo{volume}{114}},
  \bibinfo{pages}{148301} (\bibinfo{year}{2015}).

\bibitem[{\citenamefont{Weijs et~al.}(2015)\citenamefont{Weijs, Jeanneret,
  Dreyfus, and Bartolo}}]{WeijsPRL2015}
\bibinfo{author}{\bibfnamefont{J.~H.} \bibnamefont{Weijs}},
  \bibinfo{author}{\bibfnamefont{R.}~\bibnamefont{Jeanneret}},
  \bibinfo{author}{\bibfnamefont{R.}~\bibnamefont{Dreyfus}}, \bibnamefont{and}
  \bibinfo{author}{\bibfnamefont{D.}~\bibnamefont{Bartolo}},
  \bibinfo{journal}{Phys. Rev. Lett.} \textbf{\bibinfo{volume}{115}},
  \bibinfo{pages}{108301} (\bibinfo{year}{2015}).

\bibitem[{\citenamefont{Jack et~al.}(2015)\citenamefont{Jack, Thompson, and
  Sollich}}]{JackPRL2015}
\bibinfo{author}{\bibfnamefont{R.~L.} \bibnamefont{Jack}},
  \bibinfo{author}{\bibfnamefont{I.~R.} \bibnamefont{Thompson}},
  \bibnamefont{and} \bibinfo{author}{\bibfnamefont{P.}~\bibnamefont{Sollich}},
  \bibinfo{journal}{Phys. Rev. Lett.} \textbf{\bibinfo{volume}{114}},
  \bibinfo{pages}{060601} (\bibinfo{year}{2015}).

\bibitem[{\citenamefont{Hopkins et~al.}(2012)\citenamefont{Hopkins, Stillinger,
  and Torquato}}]{HopkinsPRE2012}
\bibinfo{author}{\bibfnamefont{A.~B.} \bibnamefont{Hopkins}},
  \bibinfo{author}{\bibfnamefont{F.~H.} \bibnamefont{Stillinger}},
  \bibnamefont{and} \bibinfo{author}{\bibfnamefont{S.}~\bibnamefont{Torquato}},
  \bibinfo{journal}{Phys. Rev. E} \textbf{\bibinfo{volume}{86}},
  \bibinfo{pages}{021505} (\bibinfo{year}{2012}).

\bibitem[{\citenamefont{Lu and Torquato}(1990)}]{LuTorquatoJCP90}
\bibinfo{author}{\bibfnamefont{B.~L.} \bibnamefont{Lu}} \bibnamefont{and}
  \bibinfo{author}{\bibfnamefont{S.}~\bibnamefont{Torquato}},
  \bibinfo{journal}{J. Chem. Phys.} \textbf{\bibinfo{volume}{93}},
  \bibinfo{pages}{3452} (\bibinfo{year}{1990}).

\bibitem[{\citenamefont{Quintanilla and
  Torquato}(1997)}]{QuintanillaTorquatoJCP97}
\bibinfo{author}{\bibfnamefont{J.}~\bibnamefont{Quintanilla}} \bibnamefont{and}
  \bibinfo{author}{\bibfnamefont{S.}~\bibnamefont{Torquato}},
  \bibinfo{journal}{J. Chem. Phys.} \textbf{\bibinfo{volume}{106}},
  \bibinfo{pages}{2741} (\bibinfo{year}{1997}).

\bibitem[{\citenamefont{Quintanilla and
  Torquato}(1999)}]{QuintanillaTorquatoJCP99}
\bibinfo{author}{\bibfnamefont{J.}~\bibnamefont{Quintanilla}} \bibnamefont{and}
  \bibinfo{author}{\bibfnamefont{S.}~\bibnamefont{Torquato}},
  \bibinfo{journal}{J. Chem. Phys.} \textbf{\bibinfo{volume}{110}},
  \bibinfo{pages}{3215} (\bibinfo{year}{1999}).

\bibitem[{bou()}]{bounds}
\bibinfo{note}{As a caveat, the variance increases above this bound if pixel
  particles aggregate to create extended particles spanning more than one
  pixel. This can be seen from the Ref.~\protect{\cite{DJDhudls}} result
  ${\sigma_\phi}^2(L)/\phi = \langle v\rangle/L^d$ at large $L$ for a Poisson
  pattern made from extended particles with $\phi$-weighted average volume
  $\langle v\rangle$. In particular, aggregation causes $\langle v\rangle$ to
  increase while $\phi$ remains constant.}

\bibitem[{DJD()}]{DJDhudls}
\bibinfo{note}{D. J. Durian, ``Hyperuniformity Disorder Length Spectroscopy for
  Extended Particles" (arXiv:1707.01524).}

\bibitem[{ATC({\natexlab{a}})}]{ATCjam}
\bibinfo{note}{A. T. Chieco, M. Zu, A. J. Liu, N. Xu, and D. J. Durian, ``The
  Uniformity of Soft Disk Configurations Above and Below Jamming" (in
  preparation)}.

\bibitem[{ATC({\natexlab{b}})}]{ATCfoam}
\bibinfo{note}{A. T. Chieco, A. E. Roth, S. Torquato, and D. J. Durian,
  ``Hyperuniformity of 2d Foam in Self-Similar Growth Regime" (in
  preparation)}.

\end{thebibliography}

\end{document}